\documentclass[doublecol]{epl2}

\usepackage{graphicx}

\title{Coping with Dating Errors in Causality Estimation}
\shorttitle{Coping with Dating Errors}

\author{D.A. Smirnov$^{1,2}$, N. Marwan$^{3}$, S.F.M. Breitenbach$^{4}$, F. Lechleitner$^{5,6}$, \and J. Kurths$^{3,2}$}
\shortauthor{D.A. Smirnov et al}

\institute{ $^1$Saratov Branch of V.A. Kotel'nikov Institute of
RadioEngineering and Electronics of the Russian Academy of
Sciences -- 38 Zelyonaya St., Saratov 410019, Russia \\
$^2$Institute of Applied Physics of the Russian Academy of
Sciences, 46 Ulyanova St., Nizhny Novgorod 603950, Russia \\
$^3$Potsdam Institute for Climate Impact Research, Telegraphenberg
A31, Potsdam 14473, Germany \\
$^{4}$Institute for Geology, Mineralogy \& Geophysics, Ruhr-Universit\"{a}t Bochum, Universit\"{a}tsstr. 150, 44801, Bochum, Germany\\
$^5$Department of Earth Sciences, ETH Zurich, Sonneggstrasse 5, 8092 Zurich, Switzerland \\
$^6$Department of Earth Sciences, Durham University, Durham, DH1 3LE, UK
}


\abstract{We consider the problem of estimating causal influences
between observed processes from time series possibly corrupted by
errors in the time variable (dating errors) which are typical in
palaeoclimatology, planetary science and astrophysics. ``Causality
ratio'' based on the Wiener -- Granger causality is proposed and
studied for a paradigmatic class of model system{s t}o
reveal conditions under which it correctly indicates
directionality of unidirectional coupling. It is argued that in
case of {\it a priori} known directionality, the causality ratio
allows a characterization of dating errors {and
observational noise}. Finally, we apply the developed approach to
palaeoclimatic data and quantify the influence of solar activity
on tropical Atlantic climate dynamics over the last two millennia.
A stronger solar influence in the first millennium A.D. is
inferred. The results also suggest a dating error of about 20
years in the solar proxy time series over the same period.}

\begin{document}

\maketitle

\section{Introduction}

Revealing cause-and-effect relationships between observed
processes at various time scales is an important step in
understanding many physical, biological, physiological and
geophysical systems~\cite{Pereda2005, Winterhalder2006,
Hlavackova2007, Bezruchko2010, Wibral2011, Attanasio2013,
Webber2015, Mueller2016}. Frequently, this issue must be addressed
with rather limited knowledge about the systems under study,
amounts of observational data, and dating accuracy. A general
approach to detect and quantify causal couplings, i.e., to find
out ``who drives whom'', is the Wiener -- Granger (WG)
causality~\cite{Wiener1956, Granger1969}. In its simplest version,
the idea is to check whether a present value of one process ($X$)
can be predicted more accurately using the past of a second
process ($Y$) in comparison with predictions based solely on the
past of $X$. In fact, this concept generalizes a conditional
(partial) cross-correlation~\cite{Runge2014} and has been followed
by a number of elaborations such as information-theoretic
measures~\cite{Schreiber2000, Hlavackova2007, Ay2008, Lizier2010,
SanLiang2014} and various nonlinear
approximations~\cite{Montalto2015}. Despite some limitations and
obstacles~\cite{Nalatore2007, Hahs2011, Smirnov2012, Smirnov2013},
the WG causality appears quite useful in practice, allowing
meaningful dynamical interpretations~\cite{Smirnov2014,
Smirnov2015} and becoming increasingly widely used in different
fields, such as biomedicine~\cite{Pereda2005, Wibral2011,
Mueller2016} and geophysics~\cite{Attanasio2013}.

Causal coupling estimation is also of great value in climate
science, where temporal changes of climatically sensitive
proxies~\cite{Jones2009} are the main source of information about
past climate dynamics over long time intervals. The stalagmite
YOK-I from the Yok Balum Cave in Southern Belize is especially
well dated~\cite{Ridley2015} and provides a high-resolution
reconstruction of low-latitudinal Atlantic moisture
variations~\cite{Kennett2012}. Making use of solar irradiance
reconstructions (e.g.~\cite{Steinhilber2009}), one can ask ``How
do variations in solar activity affect regional Atlantic
climate?''. Answering this question helps further delineating the
time-variant processes that drive climate variations. However,
this question leads directly to the main difficulty with such
data: dating accuracy of the reconstructions used. Uncertainties
inherent to sampling and dating methods limit our knowledge of the
time instant of each proxy observation, so that temporal ordering
of the observations from the two time series may be distorted
uniformly or irregularly in the course of time. This makes
questionable any application of the WG causality approach, which
essentially requires a clear distinction between the future and
the past.

In this Letter, we propose a solution with an appropriate
specification of the problem setting and adaptation of the WG
causality characteristics. We consider a situation where it is
known in advance that the coupling between two processes
underlying the observed time series is unidirectional, and the
problem reduces to identifying the coupling directionality.
Observational noise and dating errors may strongly affect the
results of any coupling analysis. In particular, the usual
cross-correlation function (CCF) is obviously insufficient since
even a uniform dating error moves the location of the CCF maximum
along the time axis, so that ``lead -- lag'' information is lost.
We note, however, that the WG causality approach provides two
coupling characteristics corresponding to the two directions $X
\to Y$ and $Y \to X$, which is a richer characterization than a
single CCF value. To make the WG causality work in case of dating
errors, we suggest its modification involving the definition of
the {\it causality ratio} $r_{Y \to X}$ which is the ratio of
maximized time-lagged truncated WG causalities in the directions
$Y \to X$ and $X \to Y$. We argue that if a coupling indeed exists
in the direction $Y \to X$, then under certain conditions $r_{Y
\to X} > 1$, i.e., the causality ratio is an indicator of the
coupling directionality.

We study the conditions under which this causality ratio allows us
to extract information on directionality of unidirectional
coupling or, knowing the directionality, to characterize dating
errors {and observational noise} in the analyzed time
series. As for the latter task, the mentioned palaeoclimate
problem is a relevant example where coupling is unidirectional
from solar activity variations to regional climate (reflected in
proxy reconstructions), while dating errors {and
observational noise} in the proxy signals remain largely unknown.
Here, we (i) determine the causality ratio for a class of model
systems exactly, (ii) analyze statistical properties of its
estimator in numerical simulations, and (iii) apply the approach
to palaeoclimate data using the two records mentioned above to
assess their dating accuracy and quantify the time-variant
influence of solar activity on the tropical Atlantic climate.
{Further details of the method and additional results are
given in~\cite{Suppl}.}

\section{Wiener -- Granger causality}

Let ($X(t), Y(t)$) be a bivariate random process with realizations
$(x(t), y(t))$. Denote $x_n = x(t_n)$, $y_n = y(t_n)$, where $t_n
= nh$, $n \in {\bf Z}$, and $h$ is the sampling interval. Consider
the self-predictor $x^{ind}_n = E[X(t_n)|x_{n-1},x_{n-2},\dots]$
where the expectation $E[\cdot|\cdot]$ is conditioned on the
infinite past $\{x_{n-1},x_{n-2},\dots\}$. Its mean-squared error
is $\sigma^2_{X,ind} = E[(X(t_n) - x^{ind}_n)^2]$ where the
expectation is taken over all $x_n$ and all
$\{x_{n-1},x_{n-2},\dots\}$. This error is the least over all
self-predictors for $X$. The joint predictor $x^{joint}_n =
E[X(t_n)|x_{n-1},y_{n-1},x_{n-2},y_{n-2},\dots]$ gives the least
error $\sigma^2_{X,joint}$ over all joint predictors. The
prediction improvement (PI) $G_{Y \rightarrow X} =
(\sigma^2_{X,ind}-\sigma^2_{X,joint})/\sigma^2_{X,ind}$ is a
measure of WG causality in the direction $Y \rightarrow X$.
Everything is analogous for the direction $X \rightarrow Y$.

The WG idea was first realized for stationary Gaussian
processes~\cite{Granger1969}. Then, when estimating $G_{Y
\rightarrow X}$ from a finite time series $\{ x_n, y_n
\}_{n=1}^{N}$, one truncates the (conditioning) infinite pasts at
finite numbers of terms $l_X$ and $l_{XY}$ and fits univariate and
bivariate linear autoregressive models of the orders $l_X$ and
$(l_X,l_{XY})$ to the data via the ordinary least-squares
technique. In other words, one uses the predictors
$x^{ind}_{n,l_X} = E[X_n|x_{n-1},x_{n-2},\dots,x_{n-l_X}]$ and
$x^{joint}_{n,l_X,l_{XY}} = E[X_n|x_{n-1},\dots,x_{n-l_X},
y_{n-1},\dots,y_{n-l_{XY}}]$ and gets the truncated WG causality
$G^{tr}_{Y \rightarrow X}$. The latter is often a good
approximation of $G_{Y \rightarrow X}$ even at small $l_X$ and
$l_{XY}$. The model orders can be selected via the Schwarz
criterion~\cite{Schwarz1978} and statistical significance can be
checked via Fisher's $F$-test~\cite{Seber1977}.

\section{Causality ratio}

Consider a more general setting with the original processes $X_0$
and $Y_0$, whose observed versions $X$ and $Y$ are distorted along
two lines. First, due to an amplitude noise: $X(t) = X_0(t) +
\Xi(t)$ and $Y(t) = Y_0(t) + \Psi(t)$ where $\Xi(t)$ and $\Psi(t)$
are independent observational noises with variances
$\sigma_{\Xi}^2$ and $\sigma_{\Psi}^2$, whose discrete time
realizations $\xi_n$ and $\psi_n$ are white noises. Second, due to
time uncertainty: genuine ({\it a priori} unknown) observation
instants $t_n^X$ and $t_n^Y$ deviate from the supposed regular
equidistant series $t_n = nh$: $x_n = x(t^X_n) + \xi_n$ and $y_n =
y(t^Y_n) + \psi_n$ with $t^X_n + \delta^X_n = n h$ and $t^Y_n +
\delta^Y_n= n h$, where $\delta^X_n$ and $\delta^Y_n$ stay for the
time axis (i.e. dating) errors. The latter may be rapidly
fluctuating or slowly varying and may be defined either as random
processes or deterministic functions of time. To account for the
dating errors and retain sensitivity to coupling, we use the
time-lagged WG causality: namely, $G^{tr}_{Y \rightarrow
X}(\Delta)$ is defined as prediction improvement of $x_n$ when
using the segment {$\{y_{n-\Delta/h}, \dots,
y_{n-(l_{XY}-1)-\Delta/h}\}$}. Then, we suggest to determine its
maximum over an interval of positive and negative time lags of
some width $2\Delta_m$: $G^{tr,max}_{Y \rightarrow X} =
\displaystyle \max_{-\Delta_{m} \le \Delta \le \Delta_{m}}
G^{tr}_{Y \rightarrow X}(\Delta)$. Analogously we define
$G^{tr,max}_{X \rightarrow Y}$. Finally, the causality ratio in
the direction $Y \to X$ reads

\begin{equation}
\label{ratio} r_{Y \to X} = \displaystyle \frac{G^{tr,max}_{Y
\rightarrow X}}{G^{tr,max}_{X \rightarrow Y}}.
\end{equation}

\noindent Obviously, $r_{X \to Y}=1/r_{Y \to X}$. The value of
$\Delta_m$ should be chosen so as to exceed a maximal possible
dating error to avoid missing the maximal PIs. If, moreover, the
coupling is time-delayed, locations of the PIs maxima are shifted
along the $\Delta$-axis by the value of this delay. Hence, if one
expects a time delay, then the value of $\Delta_m$ should be
selected so as to exceed the sum of the absolute values of the
coupling delay and the dating error.

We conjecture that for unidirectional coupling $Y \to X$ and
similar individual characteristics of the processes $X$ and $Y$,
the ratio $r_{Y \to X}$ is considerably greater than unity.
However, dating errors {and observational noise} along
with estimates fluctuations due to shortness of time series may
somewhat decrease $r_{Y \to X}$, which is studied below.

\section{Model system}

Since the value of $r_{Y \to X}$ may depend on many features of
the processes under study (such as characteristic times and
sampling interval) and parameters of the estimation technique
(such as $l_X$), we need to choose a reasonably simple system and
a narrow range of the parameters for which the causality ratio can
be studied in detail. As such a testing system, we use coupled
``relaxators'' (first-order decay processes):

\begin{eqnarray}
\begin{array}{rcl} \label{Example}
dX_0/dt & = & - \alpha X_0(t) + k Y_0(t) + \zeta_X(t),\\
dY_0/dt & = & - \alpha Y_0(t) + \zeta_Y(t),
\end{array}
\end{eqnarray}
\noindent where $\alpha$ determines the characteristic relaxation
time $\tau = 1/\alpha$, $k$ is the coupling coefficient, and
$\zeta_X$ and $\zeta_Y$ are independent zero-mean white noises
with autocorrelation functions $E[\zeta_X(t_1)\zeta_X(t_2)] =
E[\zeta_Y(t_1)\zeta_Y(t_2)] = \delta(t_1-t_2)$ where $\delta$ is
Dirac's delta. Eqs.~(\ref{Example}) represent a simple, but basic
class of systems which still exhibit irregular temporal behavior
and are often encountered in different fields
(e.g.~\cite{Hasselmann1976}). The squared zero-lag CCF reads here
$C^2_{X_0Y_0,0} = (\beta/4)/(1+\beta/2)$ where $\beta =
k^2/\alpha^2$ is a non-dimensional coupling strength.
$C^2_{X_0Y_0,0}$ ranges from 0 (for $k = 0)$ to 0.5 (for $k \to
\infty$) and can be used to parameterize the coupling strength as
well. The sampling rate can be conveniently characterised by the
ratio $h/\tau$.

For system (\ref{Example}) it appears possible to confine
ourselves with the orders $l_X = l_{XY} = l_Y = l_{YX} = 1$. It
can be argued that $G^{tr}_{Y \to X}(\Delta)$ obtained at $l_X =
l_{XY} = 1$ is close to $G^{tr}_{Y \to X}(\Delta)$ obtained at
$l_X = \infty$ and $l_{XY} = 1$, if the sampling interval $h$ is
not too small (e.g. $\ge 0.2\tau$)~\cite{Smirnov2014}. In
numerical simulations here, we also find that the results for
$G^{tr}_{Y \to X}(\Delta)$ with $l_X = 1$ are close to those
obtained with $l_X$ selected via the Schwarz criterion (difference
of the order of $1 \%$). Similar arguments hold for $l_{XY}$.
Then, the quantity $G^{tr}_{Y \to X}(\Delta)$ can be expressed via
the autocorrelation function (ACF) $C_{XX}(h)$ and the CCF
$C_{XY}(\Delta)$ and $C_{XY}(\Delta -
h)$~{\cite{Smirnov2013, Suppl}}. Having found ACFs and
CCF analytically, we compute the time-lagged truncated WG
causalities versus $\Delta$ and select their maxima to calculate
the causality ratio. Such a precise analysis is performed for
various coupling coefficient values, sampling intervals,
observational noise and dating error levels, while statistical
properties of the causality ratio estimator are investigated in
numerical simulations. We check if indeed $r_{Y \rightarrow X} >
1$ and assess how small $r_{Y \rightarrow X}$ can be at all. A
closer attention is paid to cases with $0.1 \le C^2_{XY,max} \le
0.2$ and WG causalities $0.01 \le G^{tr,max}_{Y \to X} \le 0.03$
which are reminiscent of those often observed in climate data
analysis in cases of statistically significant coupling detection
(e.g.~\cite{Mokhov2011} and the palaeoclimate example below).

\section{{Exact study of possible causality ratio values}}

{Before considering the central point of dating errors
identification, it is necessary to study the case of undistorted
observations $X = X_0$ and $Y = Y_0$. F}or the most practically
interesting situation{s} of not too sparse sampling
{(e.g. $h \le 0.2\tau$)}, $r_{Y \to X}$ is well above
unity, confidently indicating the correct coupling direction.
{Namely, $r_{Y \to X} = 1.6$ for $h = 0.2\tau$ and a
moderately strong coupling of $C^2_{X_0Y_0,0} = 0.1$. For rather
sparse samplings of $h \ge \tau$, the ratio $r_{Y \to X}$ gets
close to unity and, hence, cannot reliably reveal coupling
directionality.} Th{is} is similar for any coupling
strength{: i}n particular, at $h/\tau = 0.2$ the
causality ratio remains almost constant ($r_{Y \to X} \approx
1.6$) in the wide range of $0 < C^2_{XY,0} < 0.3$. For stronger
couplings, $r_{Y \to X}$ becomes even greater, up to $\approx 3$
at $C^2_{XY,0} = 0.5${. T}hus, if the sampling is not too
sparse, $r_{Y \to X}$ correctly {detects} coupling
directionality. {More details are given in~\cite{Suppl}.}

{T}hough there can be different types of dating errors,
their basic effect can be studied on a simple example where dating
errors equal a constant temporal shift half the time (e.g. for an
older half of a palaeoclimate record where accurate dating is more
difficult) and zero otherwise. Regardless which signal is
erroneously dated, only the relative dating errors matter in
causality estimation. For definiteness, we introduce the dating
errors only into the driving signal: $\delta^Y_n = const =
\delta^Y$ half time (for $n = 1, \dots, N/2$) and $\delta^Y_n = 0$
otherwise (for $n = N/2+1, \dots, N$). The ``average CCF'' of such
a nonstationary process $(X,Y)$ can be defined as the expectation
of the sample CCF computed over the entire time span and equals an
arithmetic mean of the CCFs for the two stationary halves. The
usual WG causalities defined for the entire time span are
expressed via such an average CCF in the same way as before.
Figs.~1,a,b show that the shape of the plots for the time-lagged
WG causalities and locations of their maxima change strongly when
the dating error becomes comparable with the relaxation time
$\tau$. Then, the ``correct'' $G^{tr,max}_{Y \to X}$ decreases
almost two times as compared to zero dating error, while the
opposite $G^{tr,max}_{X \to Y}$ decreases only 1.5 times. At that,
the causality ratio becomes close to unity and may even fall down
to 0.9 for the dating error greater than $\tau$. {If a
smaller or a larger portion of a time series suffers from a
uniform dating error, then the effect of the latter on the
causality ratio and the respective distortions of the plots
$G^{tr}_{Y \to X}(\Delta)$ are weaker~\cite{Suppl}, in particular,
they vanish if the entire time series is characterized with a
uniform dating error since the causality ratio involves
maximization over temporal shifts.}

\begin{figure}
\includegraphics[width = 8.4 cm]{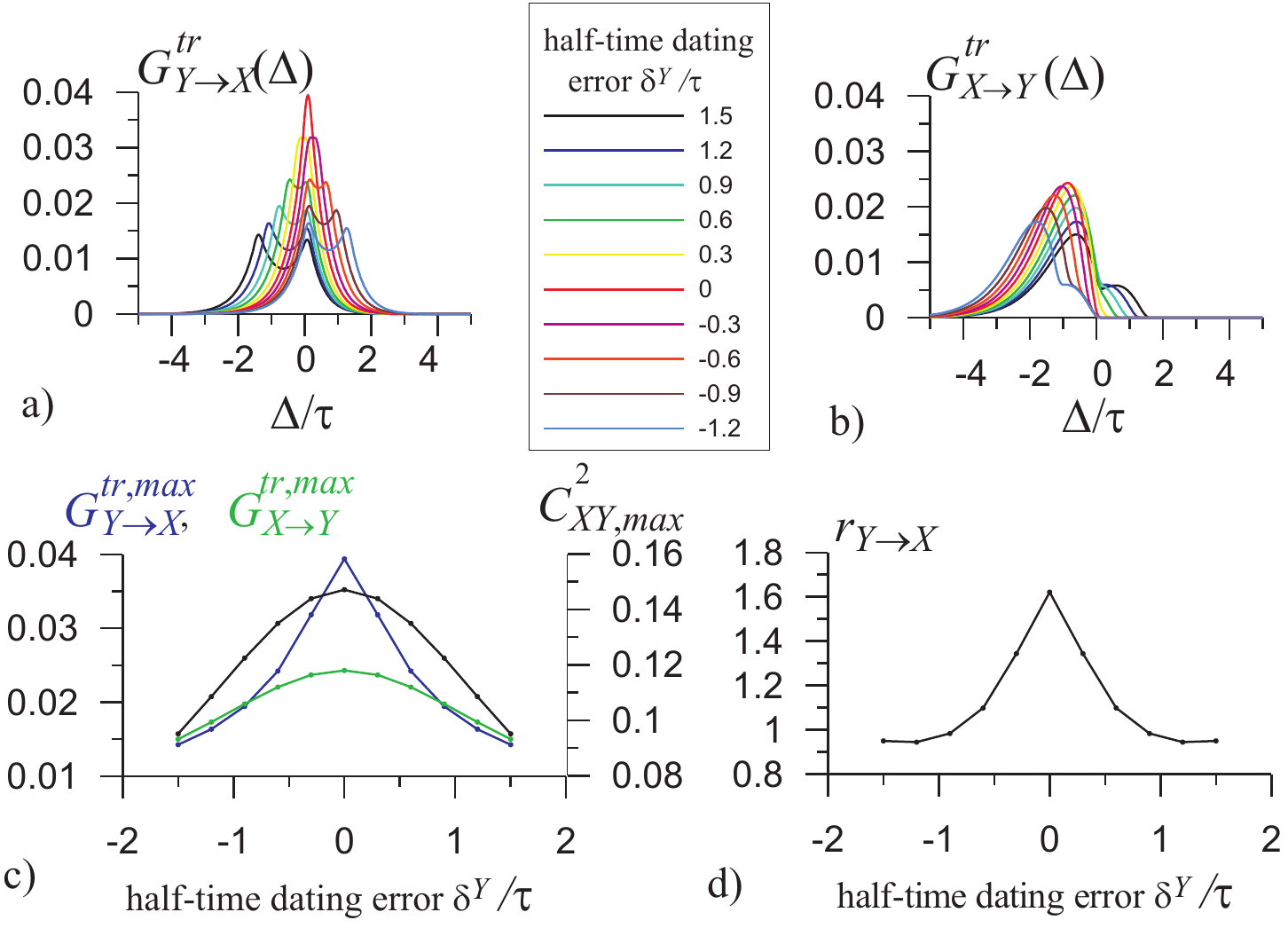}
\caption{\label{Figure1} Causality measures depending on dating
error $\delta^Y$ for the system (\ref{Example}) at $h/\tau = 0.2$
and $\sigma_{\Xi}^2 = \sigma_{\Psi}^2 = 0$, $k/\alpha$ is such
that $C^2_{X_0 Y_0,0} = 0.1$, $l_X = l_{XY} = l_Y = l_{YX} = 1$:
(a,b) truncated WG causalities versus time lag for different
dating errors; (c) maximal truncated WG causalities (blue and
green) and maximum CCF value (black) and (d) causality ratio
versus $\delta^Y$.}
\end{figure}

Principally, dating errors may be distributed in a complicated
manner determined both by random walk-like stochastic
contribution, analytical limitations and global contribution
induced by incorrect tie points as, e.g., erroneous attribution of
volcanic eruption dates due to incorrect identification of
individual eruptions~\cite{Sigl2015}. {Still}, we have
obtained results very similar {to Fig.~1} for dating
errors linearly increasing with age{, even with} a
superimposed random-walk component whose values become of the
order of $\tau$ for ages of the order of $100\tau$ as motivated by
palaeoclimate applications. Thus, the described effect of the
dating errors is robust, being observed just for reasonably large
dating errors without any other, specific conditions.

\begin{figure}
\includegraphics[width = 8.4 cm]{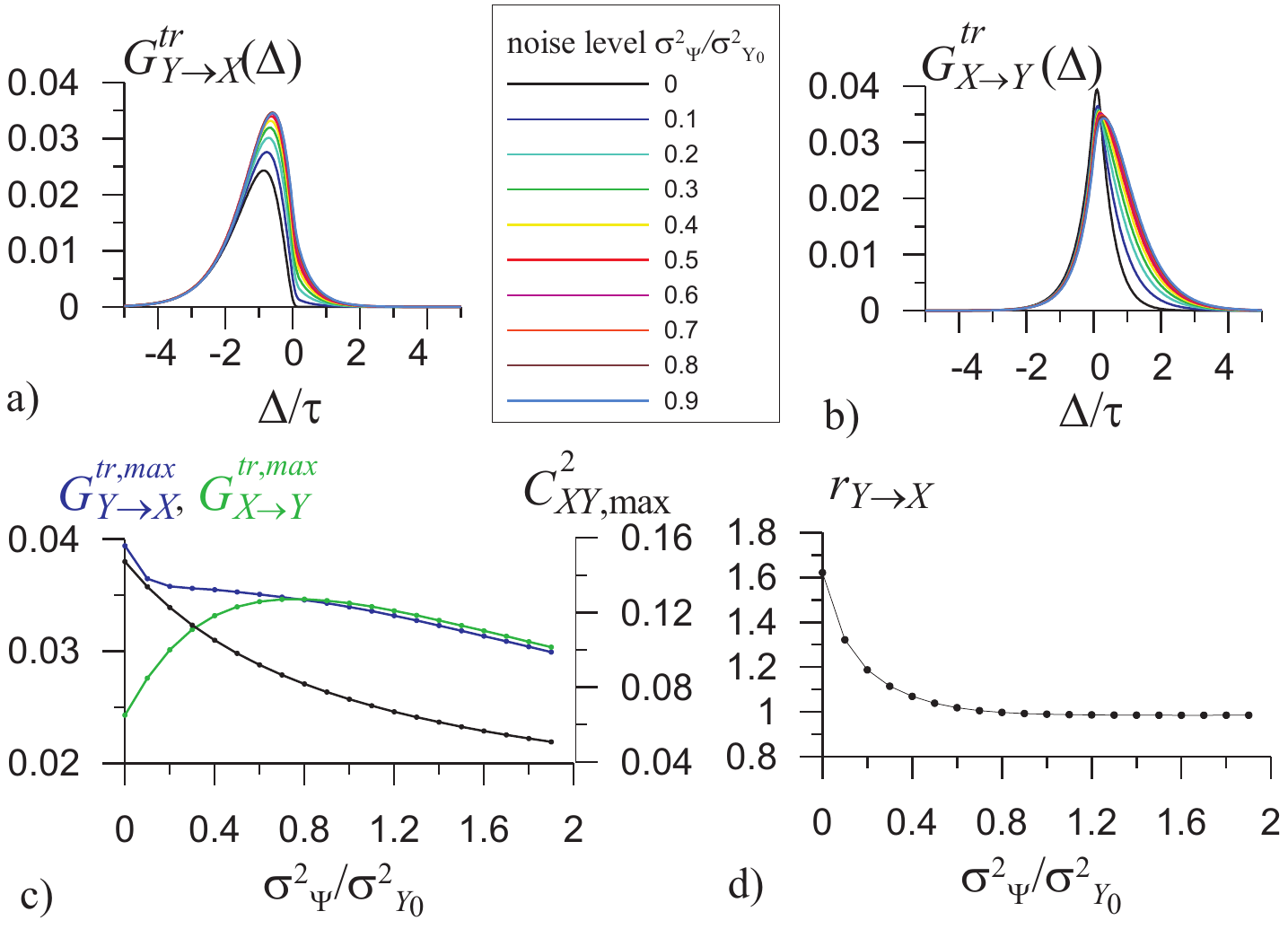}
\caption{\label{Figure2} {Causality measures depending on
observational noise level $\sigma_{\Psi}^2$ for the system
(\ref{Example}) at $h/\tau = 0.2$, $\sigma_{\Xi}^2 = 0$, $\delta^Y
= 0$, $C^2_{X_0Y_0,0} = 0.1$, and $l_X = l_{XY} = l_Y = l_{YX} =
1$: (a,b) truncated WG causalities versus time lag for different
noise levels; (c) maximal truncated WG causalities (blue and
green) and maximal CCF value (black) and (d) causality ratio
versus $\sigma_{\Psi}^2/\sigma_{Y_0}^2$.}}
\end{figure}

{When dating errors are present, it is natural to expect
also an observational noise. Let us first show how the latter
affects the causality ratio for zero dating errors. It appears}
that the noise $\Psi$ in the driving signal can significantly
decrease $r_{Y \to X}$. Thus, at moderate $h/\tau = 0.2$,
$C^2_{X_0Y_0,0} = 0.1$ and $\sigma^2_{\Xi} = 0$, the ``correct''
$G^{tr,max}_{Y \to X}$ decreases with $\sigma^2_{\Psi}$ faster
than $G^{tr,max}_{X \to Y}$ so that $r_{Y \to X}$ approaches unity
at $\sigma_{\Psi}^2/\sigma_{Y_0}^2 > 0.5$~{(Fig.~\ref{Figure2})}.
{H}owever, the noise {$\Xi$} in the driven
signal increases $r_{Y \to X}$ apart from unity, which becomes
quite visible as soon as $\sigma^2_{\Xi}/\sigma_{X_0}^2$ exceeds
{just $0.1$}. To summarize, large values of
$\sigma^2_{\Psi}/\sigma_{Y_0}^2$ ($50 \%$ and greater) along with
small $\sigma^2_{\Xi}/\sigma_{X_0}^2$ (less than $10 \%$) at
moderate coupling strengths make the ratio $r_{Y \to X}$ close to
unity{. Hence,} such a specific combination of noise
levels can complicate inference of coupling direction from $r_{Y
\to X}$.

{To distinguish between impacts of observational noise
and dating error from data, we can use either (i) assumptions
about possible levels of both factors or (ii) shapes of the plots
$G^{tr}(\Delta)$. For example, (i) if the noise is hardly greater
than $20 \%$ in terms of variance, then $r_{Y \to X} < 1.1$ may be
induced only by a dating error greater than $\tau/2$
(Figs.~2,c,d); (ii) if shapes of the plots $G^{tr}_{Y \to
X}(\Delta)$ and $G^{tr}_{X \to Y}(\Delta)$ strongly differ from
each other (cf. Figs.~1,a,b and 2,a,b), this is a sign of dating
errors rather than observational noise. Being based on exact
values of the causality ratio, such considerations are valid only
for long enough time series, where statistical fluctuations can be
neglected.}

\section{{Note of causality ratio estimation}}

{M}uch smaller causality ratio estimates (e.g., $0.5$)
could appear in practice either due to a violation of
Eq.~(\ref{Example}) or too short time series. To give an analytic
guess for possible statistical fluctuations of time series-based
estimates, we {note that the estimator $(N/l_{XY})\hat
G^{tr}_{Y \to X}(\Delta)$ for sufficiently large $N$ roughly
follows $\chi^2$ distribution with $l_{XY}$ degrees of freedom, so
the amplitude of its deviations from the mean for $l_{XY}=1$
equals} $3/N$ (the latter is the distance from $0.95$-quantile to
the mean)~\cite{Seber1977}. After maximization over a reasonable
interval of the width $2\Delta_m = 4\tau$, the difference $\delta
\hat G = \hat G^{tr,max}_{Y \to X}(\Delta)-\hat G^{tr,max}_{X \to
Y}(\Delta)$ for $l_{XY}=l_{YX}=1$ fluctuates with an amplitude of
$(3/N)\sqrt{4\cdot2} \approx 9/N$. Denote the expectation of this
difference $\delta G$. Then, $\delta \hat G$ and hence the
estimator $\hat r_{Y \to X}  = \hat G^{tr,max}_{Y \to X}/\hat
G^{tr,max}_{X \to Y}$ are slightly affected by statistical
fluctuations if the time series length is $N \gg 9/\delta G$.
Hence, for a typical $\delta \hat G \approx 0.01$ {(as in
the following example)} one should require $N \gg 900$. If a time
series is shorter, the role of statistical fluctuations may well
appear strong. For a detailed numerical study of such small sample
effects, let us focus on situations close to the properties of the
palaeoclimate data analyzed below.

\section{Causality estimates from palaeoclimate data}

A key problem in Climate Sciences is to understand and evaluate
relative contributions of different factors to observed global and
regional climate variations over time scales on the order of
decades and longer. The best sources of such information from the
pre-instrumental era are palaeoclimate proxies from different
natural archives. One well-dated high-resolution reconstruction
has been extracted from the stalagmite YOK-I from Yok Balum Cave
(Southern Belize)~\cite{Kennett2012}. The $\delta^{18}O$ record
represents local {to regional} hydroclimate variations in
{that} Atlantic region over the last two millennia with
{a} mean temporal resolution of half a year and is
characterized by very low dating errors (up to 17 yrs for ages
about 2000 yrs). This time series ($x$ signal) is examined here in
parallel with the reconstruction of the total solar irradiance
(TSI) based on $^{10}$Be measurements on ice
cores~\cite{Steinhilber2009} to extract information on a possible
influence of solar activity ($y$ signal) on the Belize climate
over the last two millennia.

The time series are presented in Figs.~3,a,b. The TSI data
(Fig.~3,b) have originally been processed to remove the 11-yr
solar cycle~\cite{Steinhilber2009} and sampled in steps of $h = 5$
yrs. The original, nonequidistantly sampled YOK-I $\delta^{18}O$
values are shown as red dots in Fig.~3,a, the blue line shows the
Gaussian kernel-based filtered\cite{Rehfeld2013} record (efficient
width of 5 yrs) sampled equidistantly in smaller steps of 1 yr.
The sample ACFs of both signals (Fig.~3,c) and their CCF
(Fig.~3,d, $\hat C^2_{XY,max} = 0.09$) agree reasonably well with
the hypothesis of the relaxators (\ref{Example}) with $\tau
\approx 25$ yrs; some deviations may be attributed to statistical
fluctuations. The resulting time series length is $N = 400$: the
signal duration is $80\tau$, the sampling interval is $0.2\tau$.

\begin{figure}
\includegraphics[width = 8.4 cm]{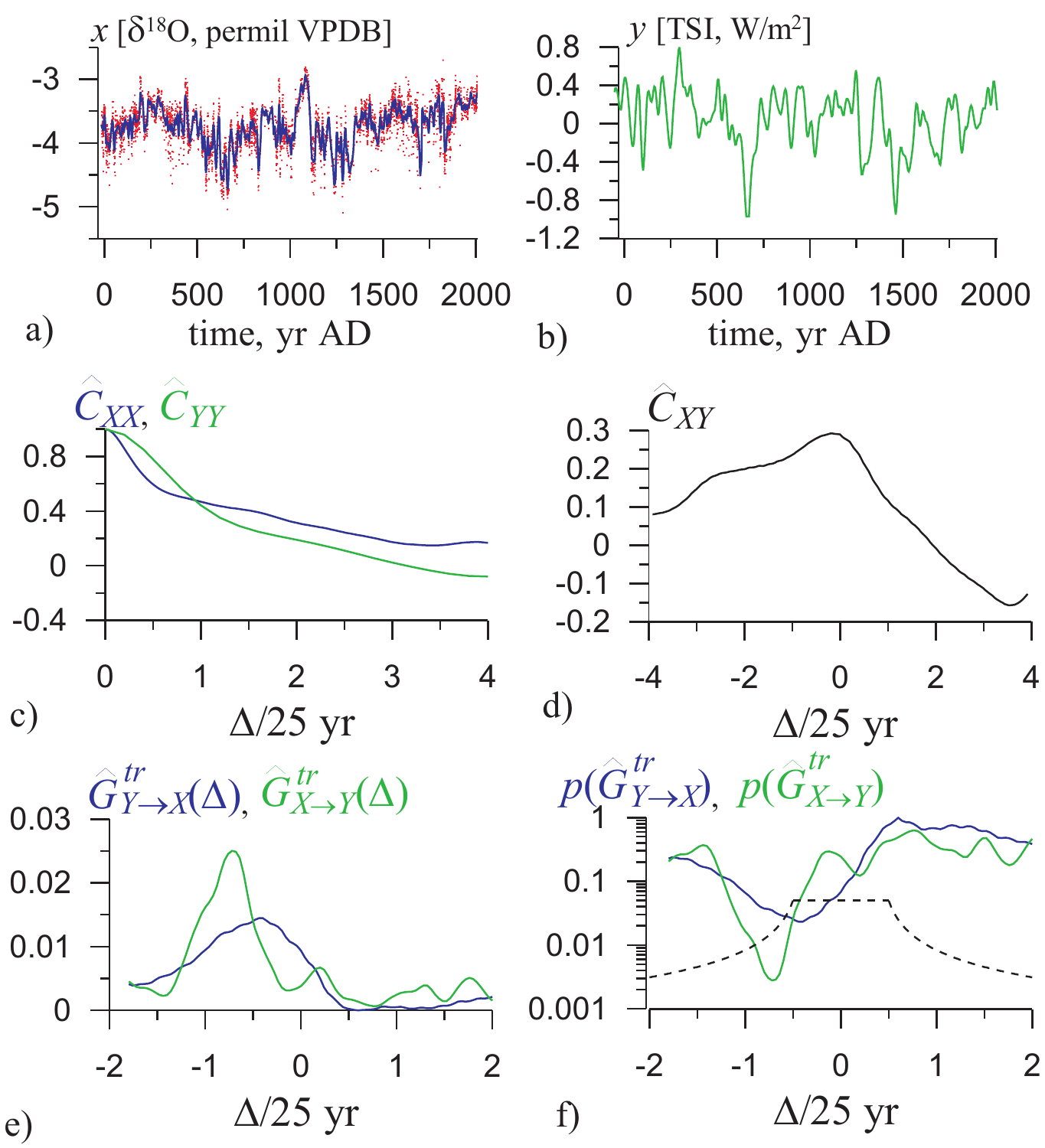}
\caption{\label{Figure3} Estimation from palaeoclimate data over
the period [15 yr BC - 2010 yr AD]: (a) time series of
$\delta^{18}O$ from a speleothem representing local climate
(moisture) in the Atlantic region, red points denote the original
data, blue line -- smoothed signal; (b) time series of solar
activity (total solar irradiance); (c) sample ACF for the signals
$x$ (blue) and $y$ (green); (d) sample CCF; (e) truncated WG
causalities in the directions TSI $\to$ Belize climate (blue) and
Belize climate $\to$ TSI (green) for $l_X = 3$, $l_{XY} = 1$, $l_Y
= 4$, $l_{YX} = 1$; (f) the respective pointwise $p$-levels for
the positivity of $\hat G^{tr}_{Y \to X}$ (blue) and $\hat
G^{tr}_{X \to Y}$ (green), black dashed lines show the pointwise
$p$-levels corresponding to the total $p$-level of 0.05 and
obtained via the Bonferroni correction~\cite{Lehmann1986} with a
pre-defined order of tests.}
\end{figure}

To focus on the most statistically reliable results, we use the
model orders selected via the Schwarz criterion for these data
($l_X = 3$ and $l_Y = 4$), even though everything is similar for
the unit orders. The WG causality estimates differ from zero at
least at the level of 0.05: $\hat G^{tr,max}_{Y \to X} = 0.014$
and $\hat G^{tr,max}_{X \to Y} = 0.025$ (Figs.~3,e,f). Since $\hat
G^{tr}_{Y \to X}(\Delta)$ for the direction TSI $\to$ Belize
climate is maximal at negative time lag $\Delta$ instead of an
expected non-negative lag, a possible dating error can be assumed.
It is surprising that the causality ratio from TSI to Belize
climate is $\hat r_{Y \to X} = 0.56$, though we would expect much
greater $r_{Y \to X} > 1.5$  without observational noise and
dating errors {a}nd $r_{Y \to X} > 0.9$ with those
distortions (Fig{s.~1,} 2). Below, we study causality
estimators for the same time series length and other parameters
and check if statistical fluctuations suffice to explain such a
low $\hat r_{Y \to X}$.

\section{Causality estimates from short time series}

Taking $N = 400$ and $h / \tau = 0.2$, we generated an ensemble of
1000 time series by integrating Eqs.~(\ref{Example}) with the
Euler -- Maruyama technique at time step of $\tau/300$ and
imposing (or not) observational noise and dating errors. From each
time series, we estimated WG causalities and causality ratio (for
$l_X = 3$, $l_Y = 4$, $l_{XY} = l_{YX} = 1$). Then we calculated
their mean values and probabilities to exceed threshold values
equal to the respective palaeoclimate
estimates{~\cite{Suppl}}. The result is that for this
{data amount} the effect of statistical fluctuations on
the causality estimates is considerably stronger than that of
dating errors (the second place) and observational noise (the
third place).

Without observational noise and dating errors, we specify $k /
\alpha = 0.45$ which gives CCF close to the palaeoclimate
estimate. For smaller $k/\alpha$ (e.g. $\leq 0.3$) the WG
causality estimates are insignificant according to the $F$-test,
while for greater $k/\alpha$ (e.g. $\geq 0.6$) the CCF and WG
causalities estimates considerably exceed the respective
palaeoclimate values. The estimation shows that typically $\hat
r_{y \to x} > 1$. A less typical case of $r_{y \to x}<1$ (even
down to 0.7) is observed in fewer than $10 \%$ of time series in
an ensemble. Both WG causality estimates are significant at least
at $p = 0.05$ in more than $90 \%$ of the time series. Appearance
of the plots $\hat G^{tr}_{Y \to X}(\Delta)$ and $\hat G^{tr}_{X
\to Y}(\Delta)$ is similar to Figs.~3,e,f, except for the
locations of the maxima{~\cite{Suppl}}: statistically
significant $\hat G^{tr}_{Y \to X}(\Delta)$ has a maximum near
zero, not at a negative lag. However, the half-time dating error
$\delta^Y = 0.8 \tau = 20$ yrs moves the maximum of $\hat
G^{tr}_{Y \to X}(\Delta)$ to negative lags of $\Delta \approx
-\delta^Y$ which is observed in about $50 \%$ of the ensemble.
Thus, the system (\ref{Example}) with dating errors is closer to
our palaeoclimate example.

The values of $\hat r_{Y \to X}$ depend on various
factors{~\cite{Suppl}}. For zero observational noise and
zero dating errors the mean of $\hat r_{Y \to X}$ is 1.2 which is
already low enough as compared to the theoretical $r_{Y \to X} =
1.6$, i.e., statistical fluctuations of the estimate already play
the role of noise. The ratio $r_{Y \to X}$ decreases very slightly
under increasing noise in the driving signal $\sigma_{\Psi}^2$
even up to a very large $100 \%$ level (at zero noise in the
driven signal). The probability to observe values of $\hat r_{Y
\to X} \le 0.56$ rises with $\sigma_{\Psi}^2$ from 0.03 only up to
to 0.05. The estimates of $r_{Y \to X}$ appear more sensitive to
the dating error and their mean falls down to 1.1 already for
moderate $\delta^Y = -0.8 \tau$ and the probability of observing
$\hat r_{Y \to X} \le 0.56$ rises from 0.03 to 0.06 at $\delta^Y =
-0.8 \tau$ and even to 0.08 at $\delta^Y = -2 \tau$ suggesting
that the dating error is more probable to be of importance here
than the observational noise. Overall, for a time series of the
considered moderate length, statistical fluctuations are more
influential than observational noise and dating errors: the former
decrease the causality ratio from 1.6 to 1.2, as compared to the
change of the order of 0.1 induced by the dating error and 0.05 by
observational noise. Thus, the time series length seems to be the
main factor limiting the accuracy of the estimation for the
palaeoclimate data at hand. Yet, as justified above, the relative
importance of each factor depends on the time series length. In
practice, it can be checked {\it ad hoc} for a time series at hand
as is done here.

{To develop a standard test for statistical significance,
we note that under the null hypothesis of uncoupled processes the
estimator $\hat r_{Y \to X}$ resembles the ratio of two
$\chi^2$-distributed quantities with $l_{XY}$ and $l_{YX}$ degrees
of freedom. Maximization of $G^{tr}(\Delta)$ over an interval of
width $2\Delta_{m} = 4\tau$ consisting of four independent
segments corresponds to maximization of $\chi^2$-distributed
quantity over four independent trials. Numerical simulations show
that for $l_{XY} = 1$ such a maximization results in the
distribution which can be approximated by the $\chi^2$ law with
two degrees of freedom. Then, $\hat r_{Y \to X}$ is distributed
according to Fisher's $F$-law with $(2,2)$ degrees of freedom.
However, quality of the approximation reduces for short time
series, where Monte-Carlo based estimation seems more reliable.}

Additional tests with simulations of a non-equidistant sampling
from (\ref{Example}) and a subsequent Gaussian kernel-based
filtering (all identical to the palaeoclimate case) show that it
slightly increases the likelihood of the causality estimates
obtained from the palaeoclimate data. Still, even in case of best
correspondence, {the} system (\ref{Example})
{exhibits} characteristics similar to those in the
palaeoclimate data only in $10 \%$ of all realizations. One reason
for this limited agreement between the data and the stationary
random process (\ref{Example}) can be temporal changes of some
characteristics of the processes underlying the proxy records.

\section{Nonstationarity of the palaeoclimate processes}

We have accounted for a possible nonstationarity  by moving window
analysis of the palaeoclimate data. The main results are presented
in Fig.~4 for two non-overlapping time windows corresponding to
the two subsequent millennia. Figs.~4,a,b (the first millennium
A.D.) reveal a usual value of the causality ratio $r_{Y \to X} =
1.05 > 1$. Figs.~4,c,d do not reveal any significant couplings for
the second millennium A.D. These results suggest a time-varying
solar effect on the Belize climate. Similar analysis with moving
windows of different lengths suggests that the transition between
the two regimes has most probably occurred over the period 1000 to
1300 A.D. A strong influence in the first millennium A.D. would be
in line with a northward position of the Intertropical Convergence
Zone (ITCZ, see also~\cite{Ridley2015}) and hence increased
rainfall in Belize. A reduced solar influence in the second
millennium A.D. could result from a southward displaced ITCZ
during the Little Ice Age, and thus reduced tropical rain in
Belize.

\begin{figure}
\includegraphics[width = 8.4 cm]{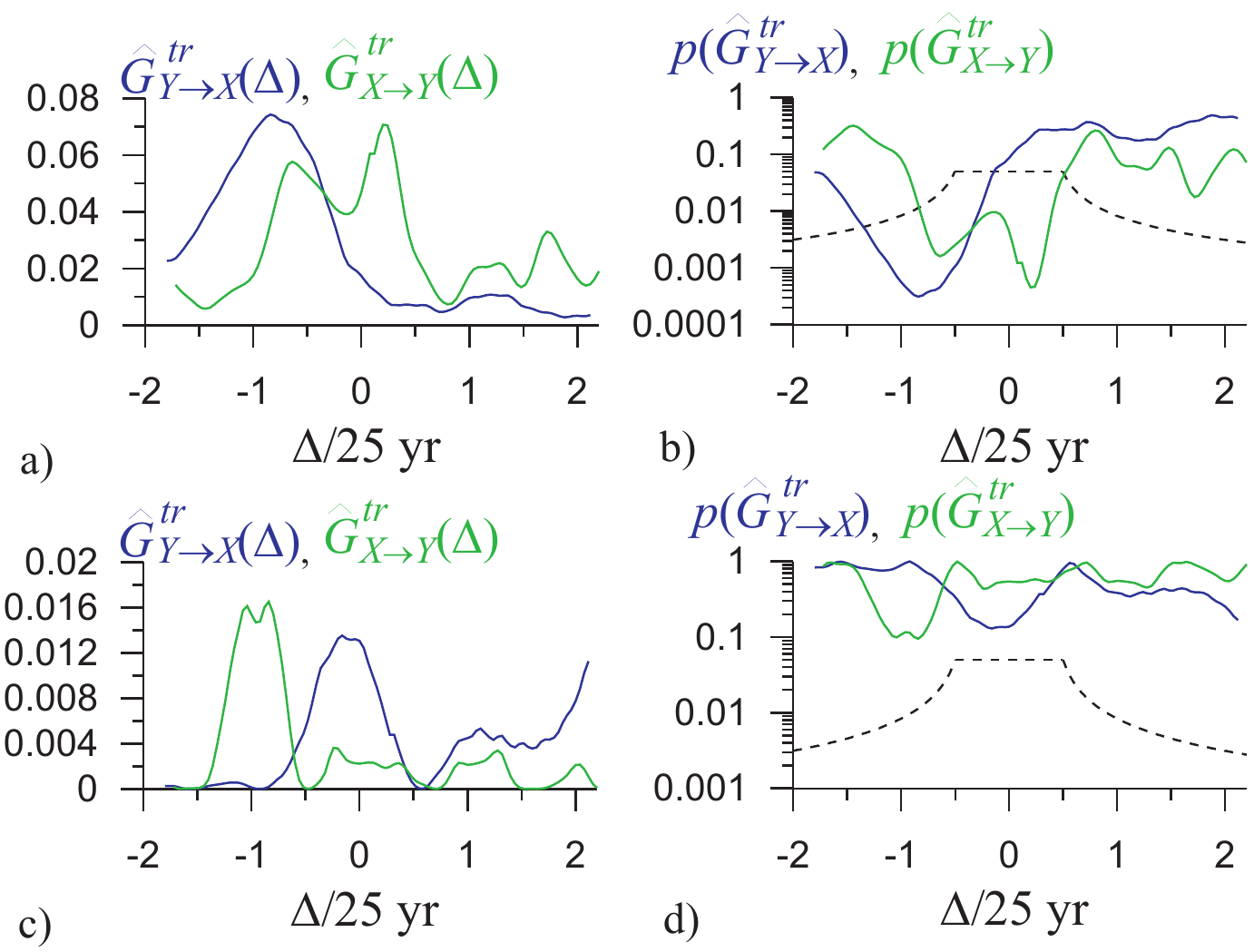}
\caption{\label{Figure4} Estimation from two non-overlapping
1000-yr intervals of the palaeoclimate data: (a,b) [15 yr BC --
985 yr AD]; (c,d) [985 yr AD -- 1985 yr AD]. Panels (a,c) show
truncated WG causalities in the directions TSI $\to$ Belize
climate (blue) and Belize climate $\to$ TSI (green) for $l_X = 3$,
$l_{XY} = 1$, $l_Y = 4$, $l_{YX} = 1$. Panels (b,d) show pointwise
p-levels for positivity of $\hat G^{tr}_{Y \to X}$ (blue) and
$\hat G^{tr}_{X \to Y}$ (green), black dashed lines show pointwise
$p$-levels corresponding to the total $p$-level of 0.05.}
\end{figure}

Our estimates for the first millennium A.D. (Figs.~4,a,b) show
that the TSI variations lag the Belize climate proxy by about 20
yrs which seems unacceptable given that TSI should always lead the
climatic signal (climatic response to the Sun). Such a lag may
well be determined by dating error of at least 20 yrs: Either the
age of the solar signal is underestimated or the age of the cave
signal is overestimated. Importantly, the question about which
signal (or both) has a larger dating error is not possible to
answer on the basis of bivariate data. We therefore include the
best-dated {ice-core based} volcanic activity
data~\cite{Sigl2015} in our analysis (instead of the TSI data) to
check whether its influence on the Belize climate (which is
expected and well-accepted) is also characterized by a
non-physical negative temporal shift~{\cite{Suppl}}. We
have found highly statistically significant volcanic forcing on
speleothem $\delta^{18}O$ variations, the maximum of $G^{tr}_{Y
\to X}(\Delta)$ being shifted to positive $\Delta = 2$ or 3 yrs,
i.e. $\Delta \approx h/2$, that agrees with the notion of volcanic
forcing delayed by no more than 1 yr. Such a small time delay is
totally acceptable. Hence, the test with the volcanic record shows
that there is excellent correspondence between eruptions
{recorded in ice cores} and YOK-I which strongly supports
the claim of highly accurate dating of the speleothem. Therefore,
we conclude that it is the TSI record which is less accurately
dated in the first millennium A.D., with a possible age
underestimation of about 20 yrs.

\section{Conclusions}

Dating errors are an almost inevitable characteristic of
palaeoclimate time series which makes causality estimation even
more difficult. We have proposed the causality ratio $r_{Y \to X}$
based on WG causality~(\ref{ratio}) as a relevant tool to cope
with this problem. We have shown that the value of $r_{Y \to X} >
1$ correctly indicates the direction of unidirectional coupling $Y
\to X$ for identical stochastic relaxators in the absence of
observational noise and dating errors, if the sampling is not too
sparse. Only very large observational noise in the driving signal
(more than $50 \%$ in terms of variance) along with the noise-free
driven signal makes $r_{Y \to X}$ close to unity and unsuitable
for coupling directionality identification. The causality ratio is
more sensitive to the dating error: if half a time series is dated
with an error about the relaxation time $\tau$ or greater, $r_{Y
\to X}$ gets close to unity again. Hence, in case of {\it a
priori} known coupling direction, the value of $r_{Y \to X}$
allows to assess likely values of dating errors and observational
noise level. However, statistical fluctuations of the estimates
from sufficiently short time series may exceed the influence of
dating errors and observational noise.

Applying the above results to analyze palaeoclimate data, we
confirmed a strong influence of solar activity on the Belize
climate over the first millennium A.D. and suggested that this
influence strongly decreased in the second millennium. An
unexpectedly low causality ratio appears to be determined by the
shortness of the time series and, probably, the dating error in
the solar proxy over the first millennium A.D. of about 20 yrs,
the age of the solar data being underestimated. It seems to be an
interesting and fruitful conclusion from an analysis of such a
short piece of data on the basis of the adapted causality
analysis.

The theoretical part of our research is based on the analysis of a
simple, but basic test system (\ref{Example}). Further studies of
the influence of dating errors and other factors on WG causalities
for more general systems are relevant, including non-identical
processes, higher dimensionality of state spaces, and various
kinds of nonlinearity. More ``inertial'' couplings can be analyzed
with $l_{XY} > 1$ and even with $l_{XY}$ different temporal shifts
rather than with a single $\Delta$. All these features will
possibly reveal more complicated relationships between the
causality ratio and coupling directionality which can then be
taken into account, extending the range of applicability of the
approach to all fields where dating errors are encountered. Yet,
the research presented here is valuable as the first step which
already reveals that the adapted WG causality analysis is a
promising tool to deal with data corrupted by dating errors and
extract information about underlying causal couplings.

\acknowledgments The work is partially supported by the Government
of Russian Federation (Agreement No. 14.Z50.31.0033 with the
Institute of Applied Physics RAS) and the European Union's Horizon
2020 Research and Innovation programme (Marie Sk\l{}odowska-Curie
grant agreement No. 691037). The theoretical and numerical study
of mathematical examples is done under the support of the Russian
Science Foundation (grant No. 14-12-00291).

\cleardoublepage

\onecolumn
\section{Supplementary Material}

\section{Definition of Wiener -- Granger causality}

Let ($X(t), Y(t)$) be a bivariate random process with $x_n = X(n
h)$, $y_n = Y(n h)$, $n \in {\bf Z}$, $h$ is sampling interval.
Self-predictor of $x_n$ given by $x^{ind}_n =
E[x_n|x_{n-1},x_{n-2},\dots]$, where $E[\cdot|\cdot]$ stands for a
conditional expectation, gives the least (over all
self-predictors) mean-squared error $\sigma^2_{x,ind} = E[(x_n -
x^{ind}_n)^2]$. The joint predictor $x^{joint}_n =
E[x_n|x_{n-1},y_{n-1},x_{n-2},y_{n-2},\dots]$ gives the error
$\sigma^2_{x,joint}$. Normalized prediction improvement value
$G_{Y \rightarrow X} =
(\sigma^2_{x,ind}-\sigma^2_{x,joint})/\sigma^2_{x,ind}$ is a
measure of WG causality in the direction $Y \rightarrow X$,
originally called ``causality strength''~\cite{Granger1963}. The
idea was suggested in Ref.~\cite{Wiener1956} and realized in
Ref.~\cite{Granger1969} in application to stationary Gaussian
process $(x_n,y_n)$. The latter yields to a bivariate linear
autoregressive (AR) equation
\begin{eqnarray}
\begin{array}{rcr} \label{BiAR2}
x_n & = & \displaystyle \sum_{k=1}^{\infty}{a_{x,k} x_{n-k}} +
\sum_{k=1}^{\infty}{b_{x,k} y_{n-k}} + \xi_{n},\\
y_n & = & \displaystyle \sum_{k=1}^{\infty}{a_{y,k} y_{n-k}} +
\sum_{k=1}^{\infty}{b_{y,k} x_{n-k}} + \psi_{n},
\end{array}
\end{eqnarray}
\noindent where $(\xi_n, \psi_n)$ is bivariate zero-mean Gaussian
white noise with variances $\sigma^2_{\xi}$, $\sigma^2_{\psi}$ and
covariance $E[\xi_n \psi_n] = \gamma$. Whiteness assures that
$\sigma^2_{\xi} = \sigma^2_{x,joint}$ and $\sigma^2_{\psi} =
\sigma^2_{y,joint}$~\cite{Box1970}. Similarly, a process $x_n$
yields to a univariate AR description, i.e. the first line of
Eqs.~(\ref{BiAR2}) with all $b_{x,k} = 0$ and white noise
$\xi^{\prime}_n$ with variance $\sigma^2_{\xi^{\prime}} =
\sigma^2_{x,ind}$. Now, $G_{Y \rightarrow X}$ can be determined.
Everything is similar for $G_{X \rightarrow Y}$.

\section{Estimation of WG causality}

In order to estimate the theoretical values $G_{y \rightarrow x}$
from a finite time series $\{ x_n, y_n \}_{n=1}^{N}$, one
truncates the infinite sums in Eq.~(\ref{BiAR2}) at finite numbers
of terms and fits truncated univariate and bivariate AR models
\begin{eqnarray}
\begin{array}{rcr} \label{TrAR2}
x_n & = & \displaystyle \sum_{k=1}^{l_X}{\tilde a_{x,k} x_{n-k}} +
\sum_{k=1}^{l_{XY}}{\tilde b_{x,k} y_{n-k}} + \tilde \xi_{n},\\
y_n & = & \displaystyle \sum_{k=1}^{l_Y}{\tilde a_{y,k} y_{n-k}} +
\sum_{k=1}^{l_{YX}}{\tilde b_{y,k} x_{n-k}} + \tilde \psi_{n},
\end{array}
\end{eqnarray}
\noindent to the data via the ordinary least-squares technique,
e.g.~\cite{Box1970}. Formally speaking, one uses the predictors
$x^{ind}_{n,l_X} = E[x_n|x_{n-1},x_{n-2},\dots,x_{n-l_X}]$ and
$x^{joint}_{n,l_X,l_{XY}} = E[x_n|x_{n-1},\dots,x_{n-l_X},
y_{n-1},\dots,y_{n-l_{XY}}]$. Thereby, one gets truncated WG
causality measure $G^{tr}_{Y \rightarrow X}$. The latter is often
a good approximation of $G_{Y \rightarrow X}$ already at quite
small values of the AR orders $l_X$ and $l_{XY}$.

The value of $l_X$ is often (and, in particular, in study of the
climate data here) selected via the Schwarz
criterion~\cite{Schwarz1978}. Namely, one minimizes the quantity
$\displaystyle \frac{N}{2}\ln{\hat \sigma^2_{\tilde
\xi}}+\frac{l_X+1}{2}\ln{N}$, where $\hat \sigma^2_{\tilde \xi}$
is the achieved mean-squared error of the one-step AR model
prediction. At any value of $l_{XY}$, the pointwise statistical
significance level $p(l_{XY})$ (probability of random error) of
the conclusion ``$G_{Y \to X} > 0$'' is checked via Fisher's
$F$-test~\cite{Seber1977}. The value of $l_{XY}$ can also be
selected via the Schwarz criterion. Alternatively, it can be
selected via minimization of the overall significance level with
the account of Bonferroni correction for multiple testing, i.e.
via minimization of the value $l_{XY}p(l_{XY})$. In our
palaeoclimate example we confine ourselves with $l_{XY} = l_{YX} =
1$ based on the Schwarz criterion. Thereby, we finally get an
estimate $\hat G_{Y \rightarrow X}$.

\section{Exact calculation of truncated WG causality}

Denote ${\bf R}({\bf z}) = \langle {\bf z}\cdot{\bf z}^{\rm T}
\rangle$ covariance matrix of a random vector ${\bf z}$, angle
brackets stand for expectation. Denote ${\bf x}_{n-1}^{l_X} =
(x_{n-1}, x_{n-2}, \dots, x_{n-l_X})^{\rm T}$ and ${\bf
y}_{n-1}^{l_{XY}} = (y_{n-1}, y_{n-2}, \dots, y_{n-l_{XY}})^{\rm
T}$, where T stands for transposition. To compute $G_{Y \to X}$,
one can use the covariance matrices ${\bf R}(x_n,{\bf
x}_{n-1}^{l_X})$, ${\bf R}({\bf x}_{n-1}^{l_X})$, ${\bf
R}(x_n,{\bf x}_{n-1}^{l_X},{\bf y}_{n-1}^{l_{XY}})$, and ${\bf
R}({\bf x}_{n-1}^{l_X},{\bf y}_{n-1}^{l_{XY}})$ of the respective
(conjugated) random vectors. These are square matrices of
dimensions $l_X+1$, $l_X$, $l_X+l_{XY}+1$, and $l_X+l_{XY}$,
respectively. According to Refs.~\cite{Barnett2009, Smirnov2013,
Hahs2013}, the truncated WG causality for stationary Gaussian
processes $x_n$ and $y_n$ relates to the determinants of these
matrices as
\begin{equation}
\label{Gaussian} G_{Y \rightarrow X}^{tr} = 1- \displaystyle
\frac{|{\bf R}(x_n,{\bf x}_{n-1}^{l_X},{\bf
y}_{n-1}^{l_{XY}})|}{|{\bf R}({\bf x}_{n-1}^{l_X},{\bf
y}_{n-1}^{l_{XY}})|}/\frac{|{\bf R}(x_n,{\bf
x}_{n-1}^{l_X})|}{|{\bf R}({\bf x}_{n-1}^{l_X})|}.
\end{equation}
\noindent If $l_X = l_{XY} = 1$, the right-hand side of
Eq.~(\ref{Gaussian}) involves only the correlations $C_{XX}(h)$,
$C_{XY}(0)$ and $C_{XY}(h)$, where correlation functions are
defined as $C_{XX}(lh)=\langle x_nx_{n-l} \rangle/\langle x_n^2
\rangle$, $C_{XY}(lh)=\langle x_ny_{n-l} \rangle/\sqrt{\langle
x_n^2 \rangle \langle y_n^2 \rangle}$, where zero mean of the
processes is taken into account and $l$ is integer .

The time-lagged WG causality $G^{tr}_{Y \rightarrow X}(\Delta)$ is
defined in full analogy with (\ref{Gaussian}) where ${\bf
y}_{n-1}^{l_{XY}}$ is replaced by ${\bf y}_{n-l}$ where $l$ is
integer and $\Delta = lh$:
\begin{equation}
\label{Gaussian2} G_{Y \rightarrow X}^{tr}(\Delta) = 1-
\displaystyle \frac{|{\bf R}(x_n,{\bf x}_n^{l_X},{\bf
y}_{n-l}^{l_{XY}})|}{|{\bf R}({\bf x}_n^{l_X},{\bf
y}_{n-l}^{l_{XY}})|}/\frac{|{\bf R}(x_n,{\bf x}_n^{l_X})|}{|{\bf
R}({\bf x}_n^{l_X})|}.
\end{equation}
\noindent If $l_X = l_{XY} = 1$, the right-hand side of
Eq.~(\ref{Gaussian2}) involves only the correlations $C_{XX}(h)$,
$C_{XY}(\Delta)$ and $C_{XY}(\Delta - h)$.

For a model system specified by stochastic differential equations
\begin{equation}
\label{model1} d{\bf z}/dt = {\bf A} \cdot {\bf z} + {\bf \xi},
\end{equation}
\noindent where ${\bf A}$ is a constant matrix and ${\bf \xi}$ is
white noise, all these covariance matrices can be found via
standard solution of linear differential equations for the second
moments~\cite{Smirnov2014}:
\begin{equation}
\label{model2} \frac {d \langle {\bf z}(0)\cdot{\bf z}(-t)^{\rm T}
\rangle}{dt} = {\bf A} \cdot \langle {\bf z}(0)\cdot{\bf
z}(-t)^{\rm T} \rangle.
\end{equation}

\section{Model system and design of numerical study}

To repeat the main text: As a model system, we consider identical
first-order decay processes
\begin{eqnarray}
\begin{array}{rcl} \label{Example}
dX_0/dt & = & - \alpha X_0(t) + k Y_0(t) + \zeta_X(t),\\
dY_0/dt & = & - \alpha Y_0(t) + \zeta_Y(t),
\end{array}
\end{eqnarray}
\noindent where $\alpha$ determines the characteristic relaxation
time $\tau = 1/\alpha$, $k$ is the coupling coefficient, and
$\zeta_X$ and $\zeta_Y$ are independent zero-mean white noises
with autocorrelation functions $\langle \zeta_X(t_1)\zeta_X(t_2)
\rangle = \langle \zeta_Y(t_1)\zeta_Y(t_2)\rangle =
\delta(t_1-t_2)$ where $\delta$ is Dirac's delta. For the system
(\ref{Example}) it appears possible to confine ourselves with the
orders $l_X = l_{XY} = l_Y = l_{YX} = 1$. The quantity $G^{tr}_{Y
\to X}(\Delta)$ at $l_X = l_{XY} = 1$ coincides exactly with
squared partial cross-correlation~\cite{Runge2014}. Since the
covariance matrices are found explicitly for the system
(\ref{Example}), we compute the time-lagged truncated WG
causalities versus $\Delta$ in the wide range $[-5\tau,5\tau]$ at
high resolution of $0.001\tau$ to select their maxima. Thereby,
the causality ratio is found at high precision.

\begin{figure}[b!]
\includegraphics[width = 9cm]{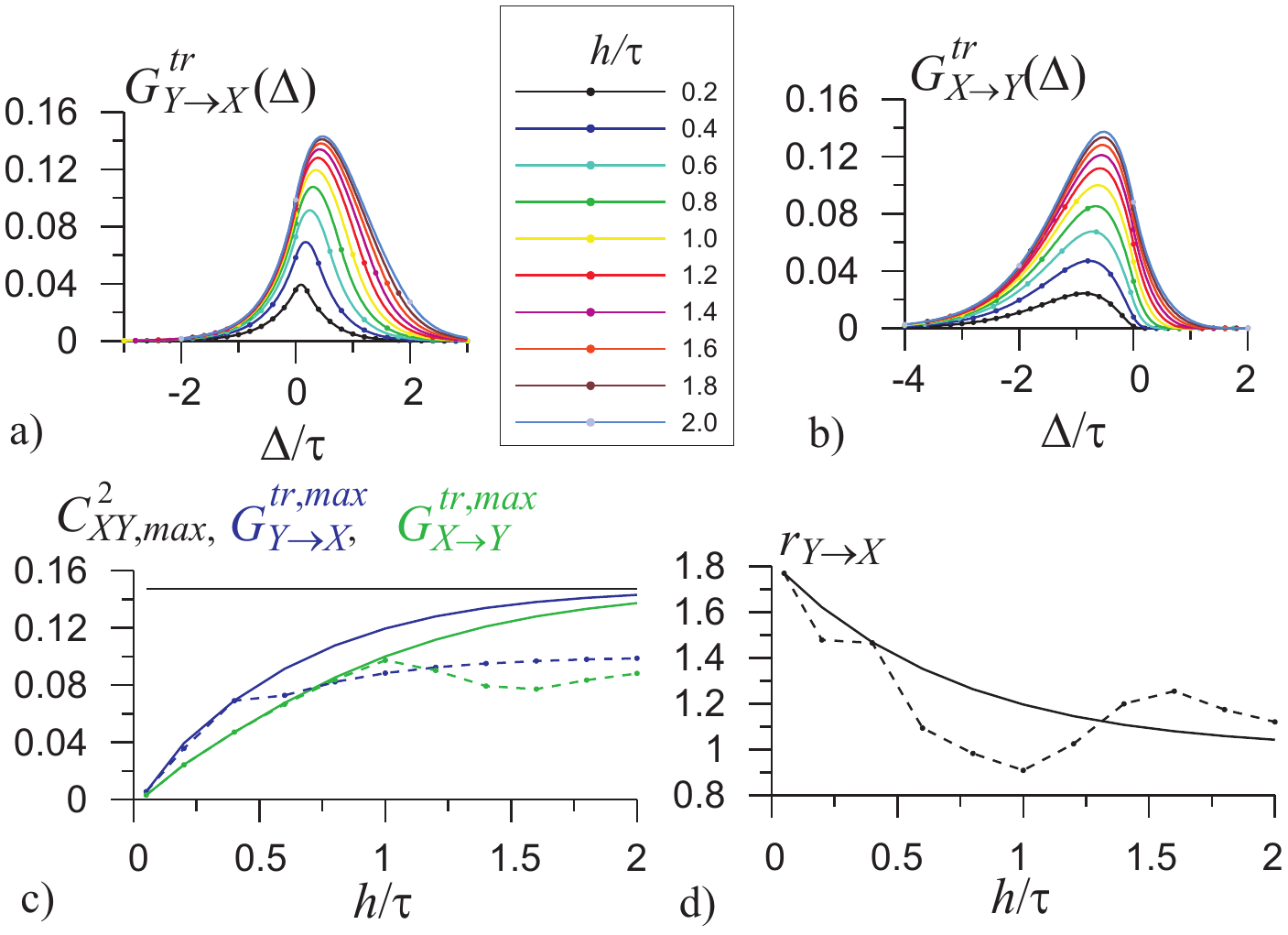}
\caption{\label{FigureS1} Causality measures for the system
(\ref{Example}) at $C^2_{X_0Y_0,0} = 0.1$ and $l_X = l_{XY} = l_Y
= l_{YX} = 1$: (a,b) truncated WG causalities versus time lag for
different sampling intervals; (c) maximal truncated WG causalities
(blue and green) and maximum CCF value (black) and (d) causality
ratio versus sampling interval. Dashed lines in (c) and (d) are
for maximization over $\Delta$ varied in steps of $h$, solid lines
-- for $\Delta$ varied in smaller steps of $0.025\tau$.}
\end{figure}

\section{Causality ratio versus sampling rate and coupling strength}

Figs.~\ref{FigureS1},a,b show $G^{tr}_{Y \to X}(\Delta)$ and $G^{tr}_{X \to
Y}(\Delta)$ for various sampling rates at moderate coupling
strength corresponding to $C^2_{XY,0} = 0.1$ ($C^2_{XY,max} =
0.15$). For a moderate $h = 0.2\tau$, the maximum value of
$G^{tr}(\Delta)$ in the ``correct'' direction $Y \to X$ is
achieved at small positive $\Delta = h/2$ (the past influences the
present), if $\Delta$ is varied at much smaller step than $h$ as
is possible when one of the signals is available at such a smaller
sampling interval (Fig.~1,a, black line), or at $\Delta = 0$, if
$\Delta$ is varied in steps of $h$ (black circles). The maximum in
the opposite direction $X \to Y$ is achieved at a ``nonphysical''
negative $\Delta = -\tau$ (Fig.~1,b) as a result of the
interdependence between $X(t)$ and $Y(t)$ induced by the $Y \to X$
coupling. This pattern of the maxima locations is characteristic
of unidirectional coupling. The causality ratio is $r_{Y \to X}
\approx 1.6$ (Fig.~\ref{FigureS1},d, solid line), which is well above unity.
Everything is similar for much smaller $h$, with $r_{Y \to X}
\approx 1.8$.

As for the rather sparse sampling with $h \ge \tau$, the ratio
$r_{Y \to X}$ gets close to unity, since $G^{tr}_{Y \to
X}(\Delta)$ and $G^{tr}_{X \to Y}(\Delta)$ become almost
independent of the conditioning variables $x(t-h)$ and $y(t-h)$
tending to the squared CCF (Fig.~\ref{FigureS1},a-c). For discrete $\Delta$,
the ratio $r_{Y \to X} \to 1$ in a non-monotone manner, taking the
values as small as 0.9 (Fig.~\ref{FigureS1},d, dashed line). Thus, only if the
sampling interval is of the order of the characteristic time
$\tau$, the causality ratio cannot reliably reveal the coupling
directionality.

\begin{figure}[t!]
\includegraphics[width = 10 cm]{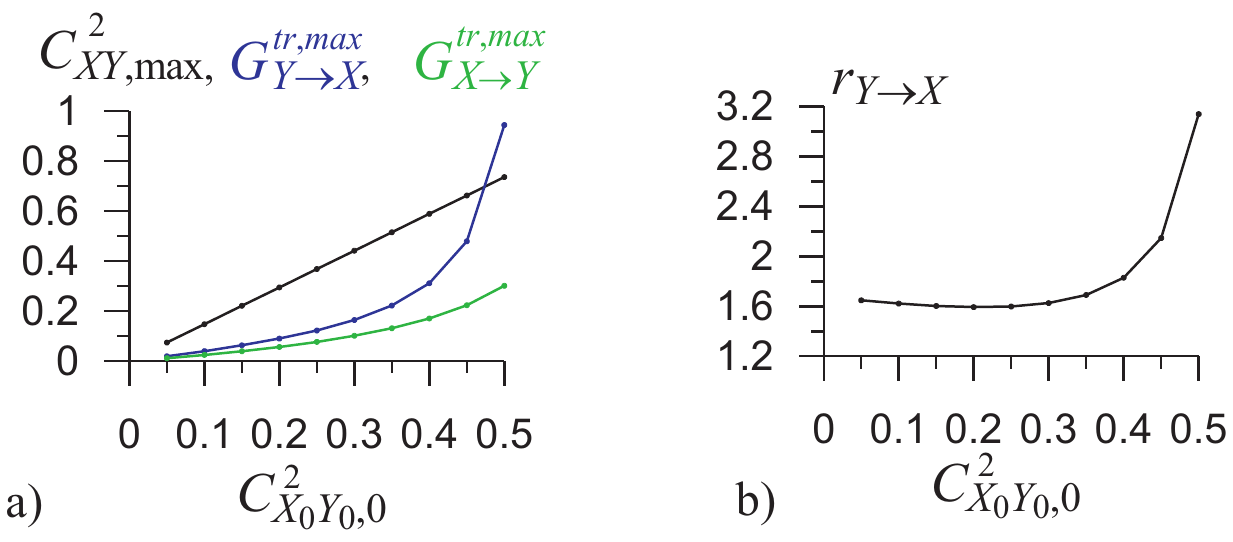}
\caption{\label{FigureS2} Causality measures for the system
(\ref{Example}) at $\Delta t/\tau = 0.2$ and $l_X = l_{XY} = l_Y =
l_{YX} = 1$ versus squared zero-lag CCF at zero observational
noise: (a) Maximal truncated WG causalities (blue and green) and
maximal CCF value (black); (d) causality ratio.}
\end{figure}

The situation is similar for any coupling strength. Fig.~\ref{FigureS2},a,b
show dependencies of the causality characteristics on
$C^2_{X_0Y_0,0}$ at $\Delta t/\tau = 0.2$. The causality ratio
achieves its maximal value of $\approx 3$ at $C^2_{X_0Y_0,0} \to
0.5$ when dynamics of the system $X$ is sustained entirely by the
system $Y$ and $C_{X_0Y_0,max} = 0.86$. The causality ratio
remains almost constant and equal to $\approx 1.6$ in the wide
range of $C^2_{X_0Y_0,0}$ from 0.05 to 0.3 (Fig.~\ref{FigureS2},b). This range
corresponds to the maximal CCF ranging from the (notable) value of
0.27 to the (rather large) value of 0.66. In particular, this
range includes the most interesting for us moderate maximal CCFs
about 0.3 -- 0.4. Thus, if the sampling is not too sparse and
cross-correlation is not too low, the causality ratio in the
``correct'' direction is considerably greater than unity (1.6 and
greater) which should allow one to confidently infer coupling
direction in practice from a sufficiently long time series.

\begin{figure}[hbtp]
\includegraphics[width = 8cm]{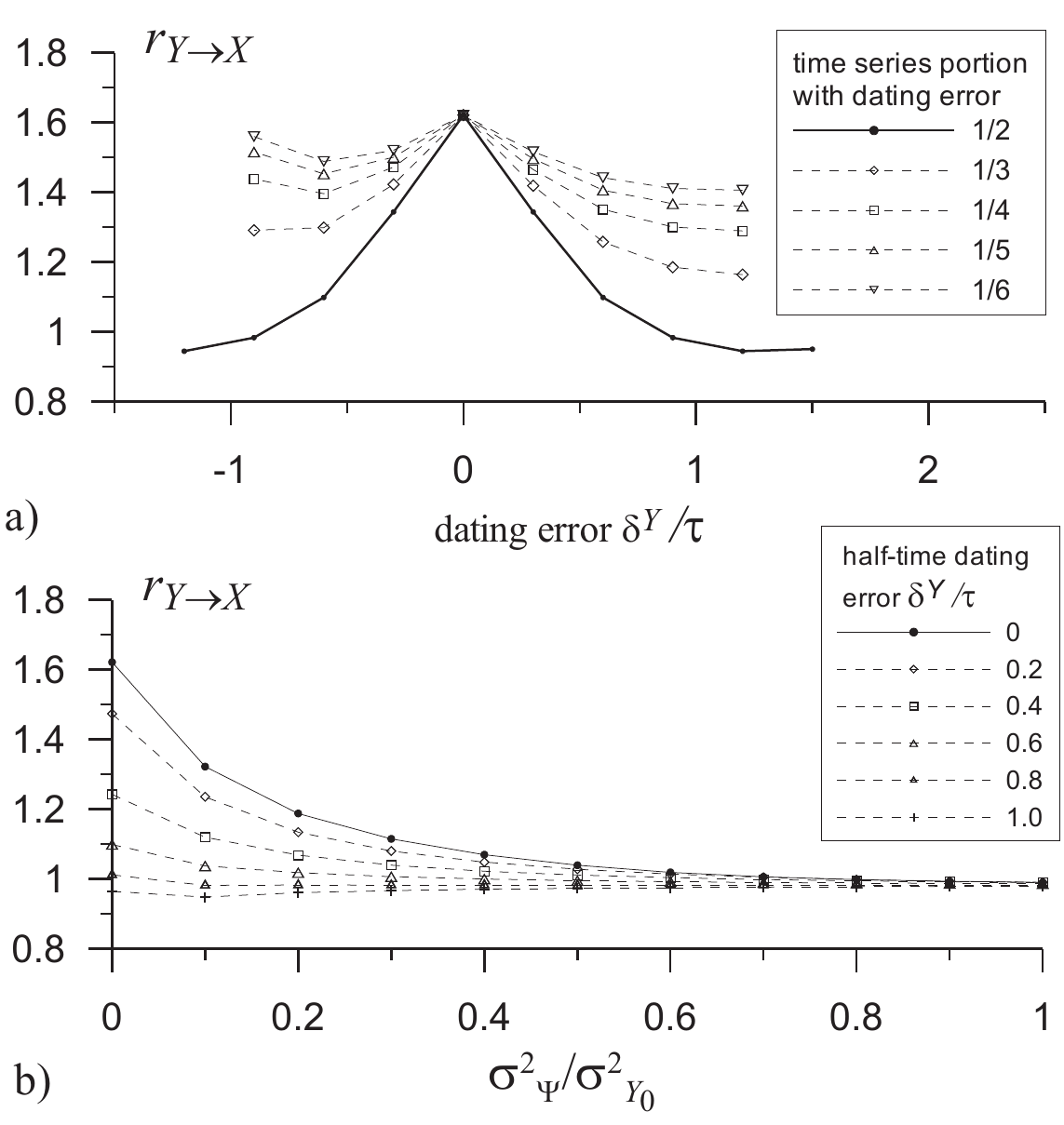}
\caption{\label{FigureS3} Causality measures for the system
(\ref{Example}) at $\Delta t/\tau = 0.2$ and $C^2_{X_0Y_0,0} =
0.1$ (a) versus dating error for zero observational noise and
different portions of the time series corrupted by the dating
error and (b) versus observational noise level in the driving
signal at for $\sigma_{\Xi}^2 = 0$ and different half-time dating
errors.}
\end{figure}

\section{Causality ratio versus dating errors and observational noise}

Fig.~\ref{FigureS3},a presents the causality ratio versus dating error for the
situation when different portion of the time series
$\{y_n\}_{n=1}^N$ (from 1/2 to 1/6 of the entire series) is
corrupted by the dating error. One can see that half-time dating
error reduces the value of the causality ratio most strongly (the
solid curve). Smaller portions distorted by the uniform dating
error lead to a weaker reduction of the causality ratio (dashed
curves). Larger uniform error-corrupted portions of 2/3, 3/4, 4/5,
and 5/6 lead to the same causality ratio reduction as the smaller
complementary ones of 1/3, 1/4, 1/5, and 1/6, respectively (not
shown in the plots). Indeed, if the entire series suffers from a
uniform dating error, this does not influence the causality ratio
since maximization over temporal shifts is involved in the
definition of the latter.

Fig.~\ref{FigureS3},b presents simultaneous influence of the half-time dating
error and observational noise in the driving signal
$\sigma_{\Psi}^2$. One can see that their contributions to the
reduction of the causality ratio $r_{y \to x}$ can sum up: e.g.
dating error of $0.2\tau$ reduces the causality ratio as compared
to zero dating error approximately by 0.15 both for
$\sigma_{\Psi}^2 = 0$ and $\sigma_{\Psi}^2 = 0.1\sigma_{Y_0}^2$,
while $\sigma_{\Psi}^2 = 0.1\sigma_{Y_0}^2$ reduces the causality
ratio as compared to $\sigma_{\Psi}^2 = 0$ approximately by 0.3
both for zero dating error and $\delta^Y = 0.2\tau$. However, for
stronger errors of both kinds their effects do not simply add: For
dating error of $0.8\tau$ and greater, the causality ratio
saturates at the level of unity and the noise does not reduce it
any more (and in some range of the noise levels it even increases
the causality ratio). Similar saturation of the causality ratio
values exists for the noise of about $\sigma_{\Psi}^2 =
0.6\sigma_{Y_0}^2$ and greater. However, the latter is a huge
noise level (about 80 \% in root-mean-squared amplitude), while
the dating error of $0.8\tau$ is quite realistic for palaeoclimate
studies, including the example considered in this work. Thus, the
capability of the dating error to decrease $r_{y \to x}$ seems to
be stronger and more robust. Still, we note that even the two
factors together cannot make $r_{y \to x}$ considerably less than
unity, only the ranges of their values leading to $r_{y \to x}
\approx 1$ widen in the presence of another factor.

\begin{figure}[htpb]
\includegraphics[width = 9 cm]{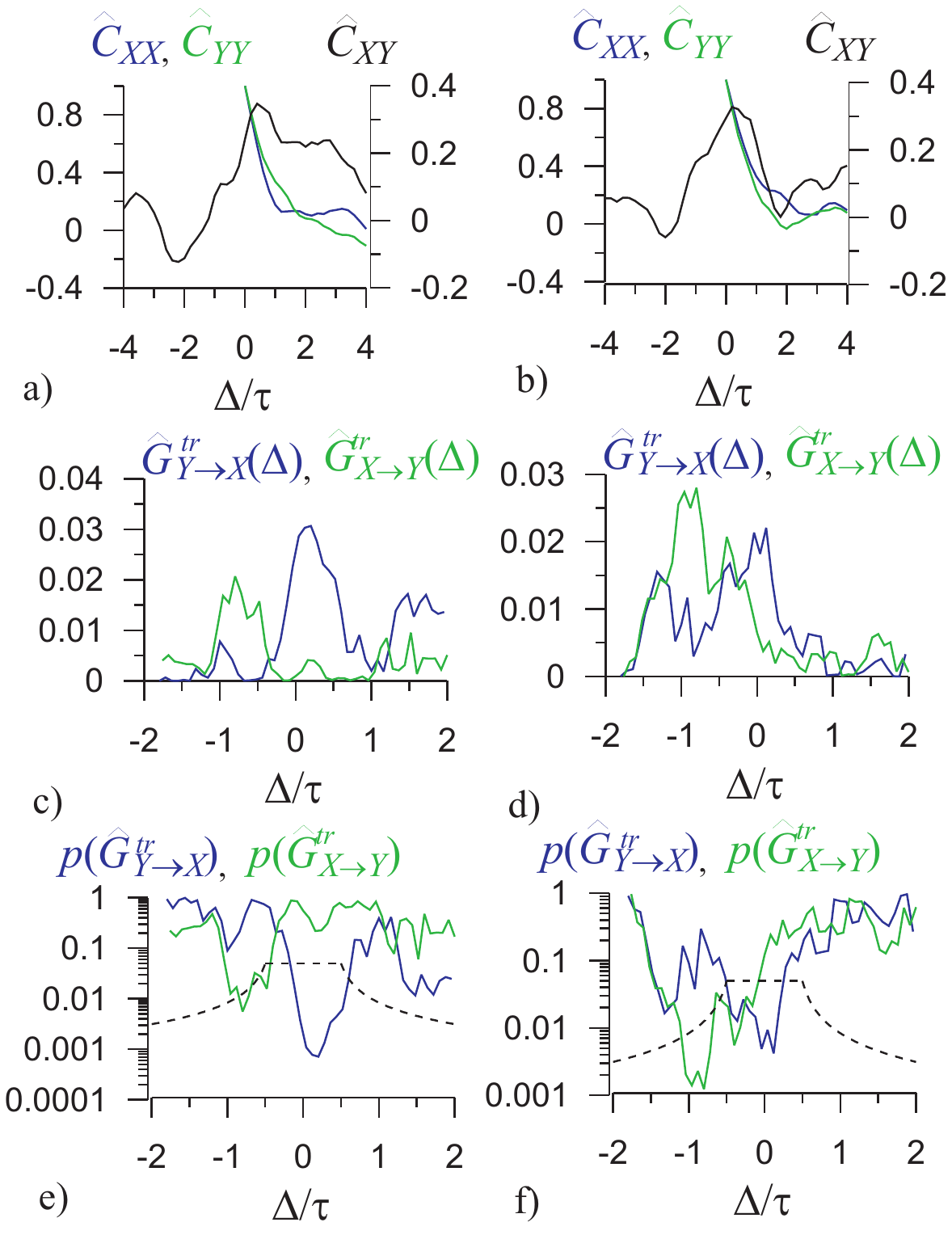}
\caption{\label{FigureS4} Two examples (left and right column,
respectively) of correlation and causality estimates from time
series of the system (\ref{Example}) versus time lag at $\Delta
t/\tau = 0.2$, $\sigma_{\Psi}^2 = 0$, $k = 0.0015$, $\delta^Y =
0$: (a,b) ACF for the signals $x$ (blue) and $y$ (green) and CCF
(black); (c,d) WG causality in the directions $Y \to X$ (blue) and
$X \to Y$ (green); (e,f) $F$-test based significance level
estimates (pointwise p-levels) for positivity of $G_{Y \to X}$
(blue) and $G_{x \to y}$ (green), black dashed lines show the
pointwise $p$-level corresponding to the global $p$-level of 0.05
(Bonferroni correction~\cite{Lehmann1986} with a pre-defined order
of tests).}
\end{figure}

\section{Causality estimates from time series: Numerical simulations}

As discussed in the main text, the causality ratio is slightly
affected by the estimator fluctuations for the estimated values of
palaeoclimate prediction improvements(of the order of 0.01), if
the time series length is $N > 900$. For the paleoclimate data at
hand we have a smaller value of $N = 400$ (the signal duration of
$80\tau$ at sampling interval $0.2\tau$) so that the role of
statistical fluctuations may well appear strong. Therefore, we
performed numerical experiments with estimation of WG causalities
and causality ratio from time series with the above parameters $N
= 400$ and $\Delta t / \tau = 0.2$ from the system (\ref{Example})
with $\alpha = 1/300$ month$^{-1} = 1/25$ yr$^{-1}$. To generate
the time series, we integrated the with Euler -- Maruyama
technique with time step of $\tau/300 = 1$ month and sampling
interval of $\Delta t = 60$ months which is analogous to the
paleoclimate data below. An ensemble of 1000 time series was
generated at each set of parameter values. Mean values of WG
causalities and causality ratio and probability of them to exceed
the respective experimentally observed paleoclimate estimates are
computed from each ensemble.

Starting with the case of absent observational noise and dating
errors, we specify $k / \alpha = 0.45$, i.e. $k = 0.0015$
month$^{-1}$ which appears overall the most close to the observed
paleoclimate data properties. In the selected case, we get mean
value of the maximal sample CCF equal to $0.33$ and the
probability for it to exceed the paleoclimate value of $0.29$
equal to $0.67$. Here, we present estimates for the truncated WG
causality $G^{tr}_{Y \to X}$ for $l_X = 4, l_{XY} = 1$ rather than
for $l_X = l_{XY} = 1$ to be consistent with the paleoclimate
example where the orders were selected via the Schwarz criterion.
However, numerical experiments show that the causality ratio
estimates in these two cases are very close to each other, in
particular, their statistics (mean values and probabilities)
differ by no more than $1 \%$. This is a further confirmation that
the above results for $l_X = l_{XY} = 1$ are correct for (and at
least qualitatively agree with) those for higher AR orders, in
particular, for the Schwarz criterion-based orders $l_X,l_{XY}$.

\begin{figure}[b!]
\includegraphics[width = 9 cm]{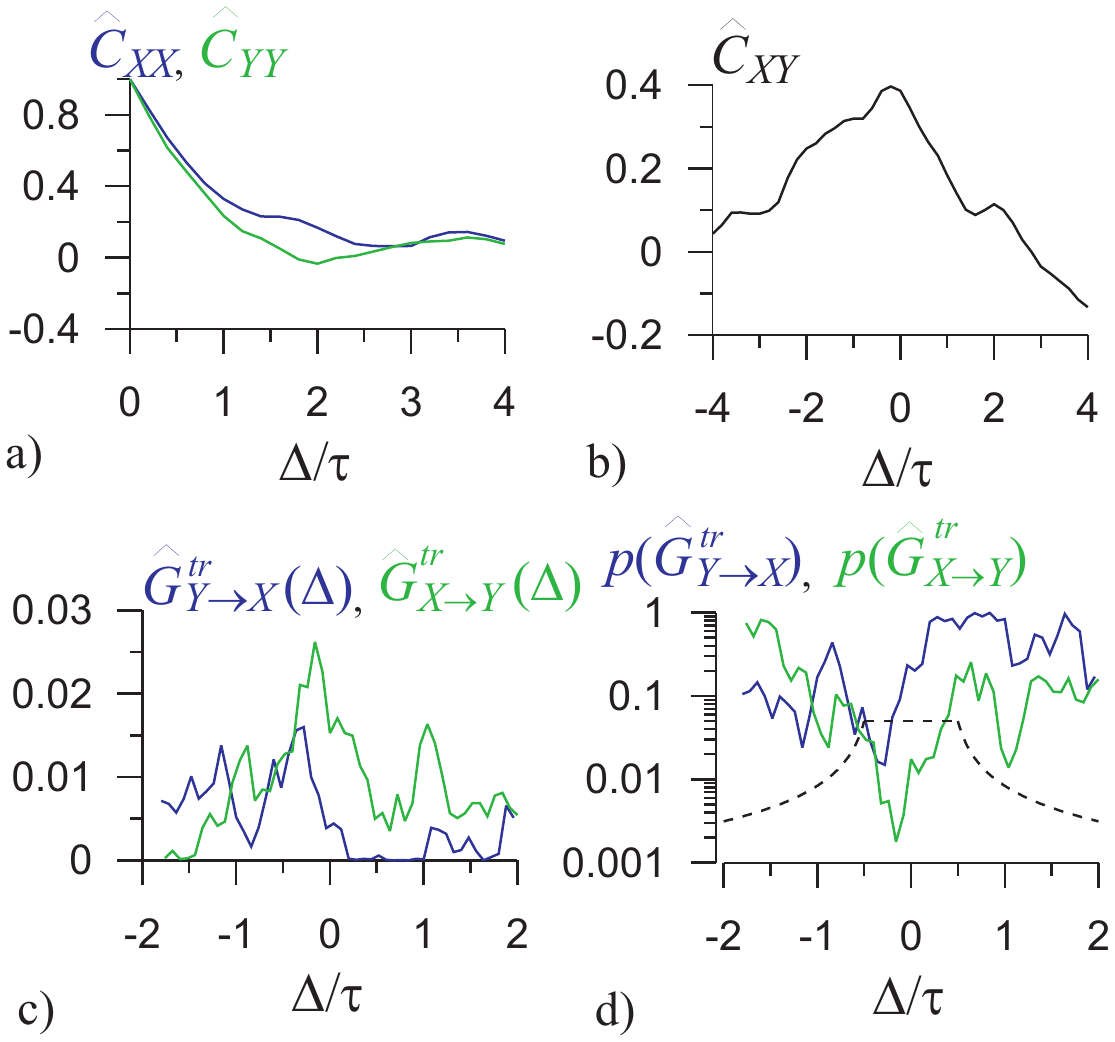}
\caption{\label{FigureS5} An example of correlation and causality
estimates from a time series of the system (\ref{Example}) versus
time lag at at $\Delta t/\tau = 0.2$, $\sigma_{\Psi}^2 = 0$, $k =
0.0015$, half-time dating error $\delta^Y/\tau = 0.8$: (a) ACF for
the signals $x$ (blue) and $y$ (green); (b) CCF; (c) WG causality
in the directions $Y \to X$ (blue) and $X \to Y$ (green); (d)
pointwise p-levels for positivity of $G_{Y \to X}$ (blue) and
$G_{X \to Y}$ (green), black dashed lines show the pointwise
$p$-level corresponding to the global $p$-level of 0.05.}
\end{figure}

Fig.~\ref{FigureS4} presents two examples of estimates obtained from two
different time series of the system (\ref{Example}): left column
is the most typical case where the causality ratio $r_{Y \to X}$
is greater than unity (namely about 1.5, Fig.~\ref{FigureS4},c), right column is
less typical case observed in less than for $10 \%$ of the time
series in the ensemble where $r_{Y \to X}<1$ (namely about 0.7,
Fig.~\ref{FigureS4},d). Both WG causality estimates are statistically
significant at least at the level of 0.05 according to the
$F$-test with Bonferroni correction, which takes place for more
than $90 \%$ of the time series in the ensemble. As for ACF and
CCF estimates they look quite similar for both cases (Fig.~\ref{FigureS4},a,b).
The right column is quantitatively similar to the paleoclimate
example except for the positions of the maxima in WG causality
plots. For the correct direction, the maximum is located close to
zero in contrast with the paleoclimate example where it is located
at a negative lag. Sometimes, the maxima for the correct direction
can appear at negative lags in this mathematical example as well,
but these cases correspond to statistically insignificant WG
causality estimates.

The half-time dating error $\delta^Y = -0.8 \tau = -20$ yr
(Fig.~\ref{FigureS5}) moves the plots for WG causality estimates along the
abscissa axis. In particular, the maximum of the plot for the
correct direction moves to the negative lags of $\Delta \approx
\delta^Y$ (Fig.~\ref{FigureS5},c,d). This location of the maxima is similar to
those for paleoclimate data and is observed in about $50 \%$ of
cases for the analyzed ensemble. Thus, we could say that the
system (\ref{Example}) with half-time dating error exhibit some
properties close to those for paleoclimate data.

\begin{figure}[t]
\includegraphics[width = 10 cm]{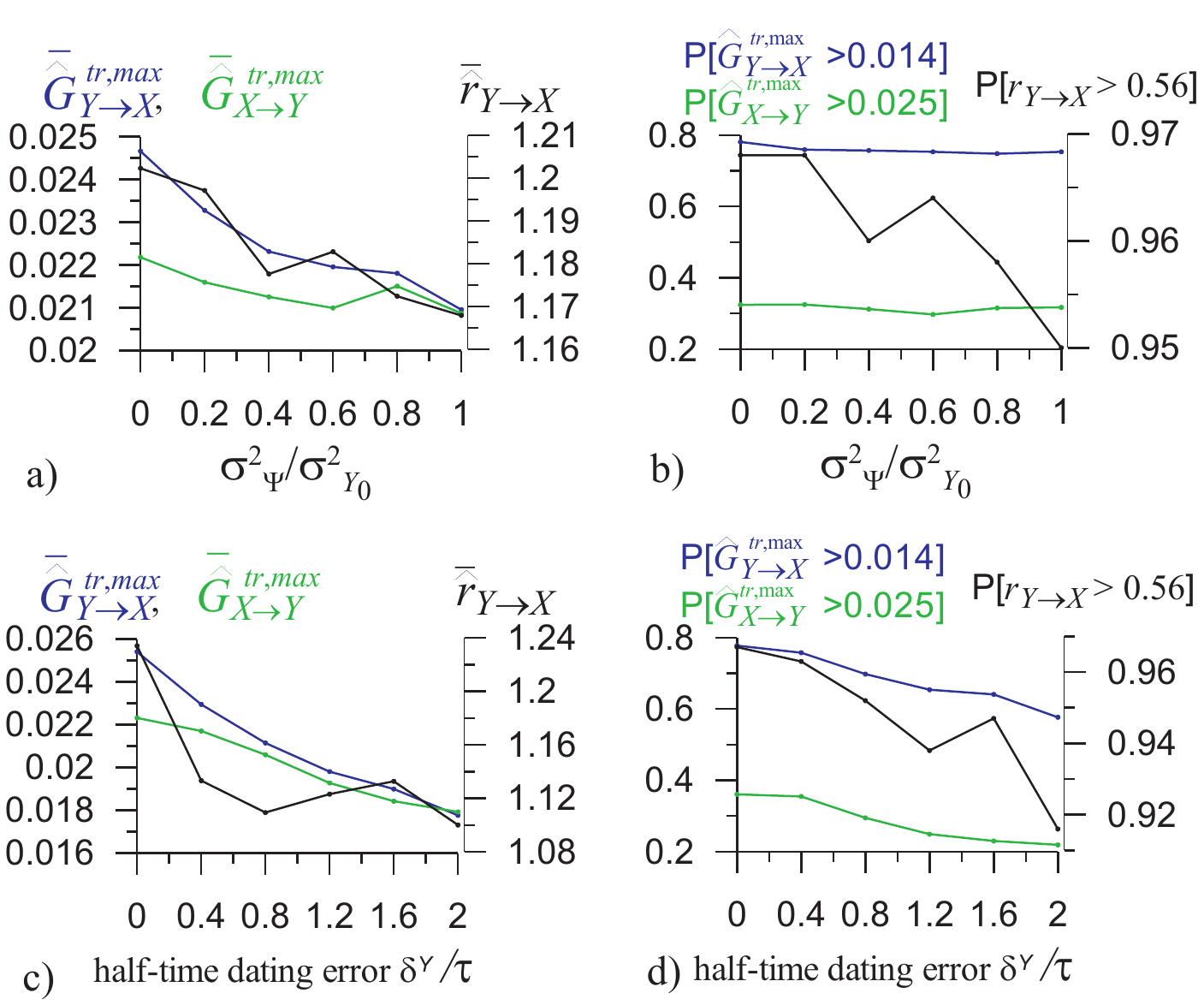}
\caption{\label{FigureS6} Statistics of the causality estimates for
the system (\ref{Example})  depending on observational noise level
at zero dating error (a,b) and on half-time dating error at zero
observational noise (c,d)  over an ensemble of 1000 time series of
the length $N = 400$ at sampling interval $\Delta t/\tau = 0.2$.
The left column shows mean values for the maximum truncated WG
causalities (blue and green) and the causality ratio (black lines,
right ordinate axes). The right column shows probabilities for
these estimates to exceed the same estimates obtained from the
paleoclimate data.}
\end{figure}

To study a dependence of the estimated causalities on noise level
and dating error, let us consider Fig.~\ref{FigureS6}. Note that the mean value
of the estimate of the causality ratio $r_{Y \to X}$ is already
low enough already for zero noise since statistical fluctuations
play the role of noise and move the estimated causality ratio
close to unity that the theoretical value (1.2 as compared to 1.6,
Fig.~\ref{FigureS6},a, black line). Fig.~\ref{FigureS6},a further shows that mean values of
WG causalities somewhat decrease with the noise level, but the
causality ratio decreases very slightly from 1.2 to 1.17 at the
very large $100 \%$ noise. As for the probabilities to exceed the
fixed ``paleoclimate'' values, Fig.~\ref{FigureS6},b shows that they are
constant for WG causalities, but for the causality ratio the
probability to observe such a low value as 0.56 rises from 0.03 to
0.05 with the noise level. Overall, the causality ratio estimates
appear weakly sensitive to observational noise, even though a very
large noise makes the observed paleoclimate estimate somewhat more
probable, prompting that the solar activity signal might be more
noise-corrupted than the Atlantic climate proxy.

As for the dating error, Figs.~\ref{FigureS6},c,d show that the causality ratio
estimates are more sensitive to this quantity. Thus, $r_{Y \to X}$
falls down to 1.1 already for moderate $\delta^Y = -0.8 \tau$ and,
more importantly, probability of observing so low causality ratio
rises from 0.03 to 0.06 at $\delta^Y = -0.8 \tau$ and even to 0.08
at $\delta^Y = -2 \tau$ making the observed paleoclimate estimate
of $r_{Y \to X}$ even more probable, prompting that the dating
error might well be present in the earlier parts of the
paleoclimate data at hand.

Overall, we must state that the contribution of statistical
fluctuations is much more important than impacts of the dating
error and observational noise. The former circumstance decreases
the causality ratio on average from 1.6 to 1.2, as compared to the
average change of the order of 0.1 induced by the dating error and
0.05 by the observational noise. Thus, the time series length
seems to be the main factor limiting the accuracy of estimation
for the paleoclimate data at hand.

\section{Estimation of causality between volcanic activity and Yok Balum speleothem-based data}

\begin{figure}[t]
\includegraphics[width = 9 cm]{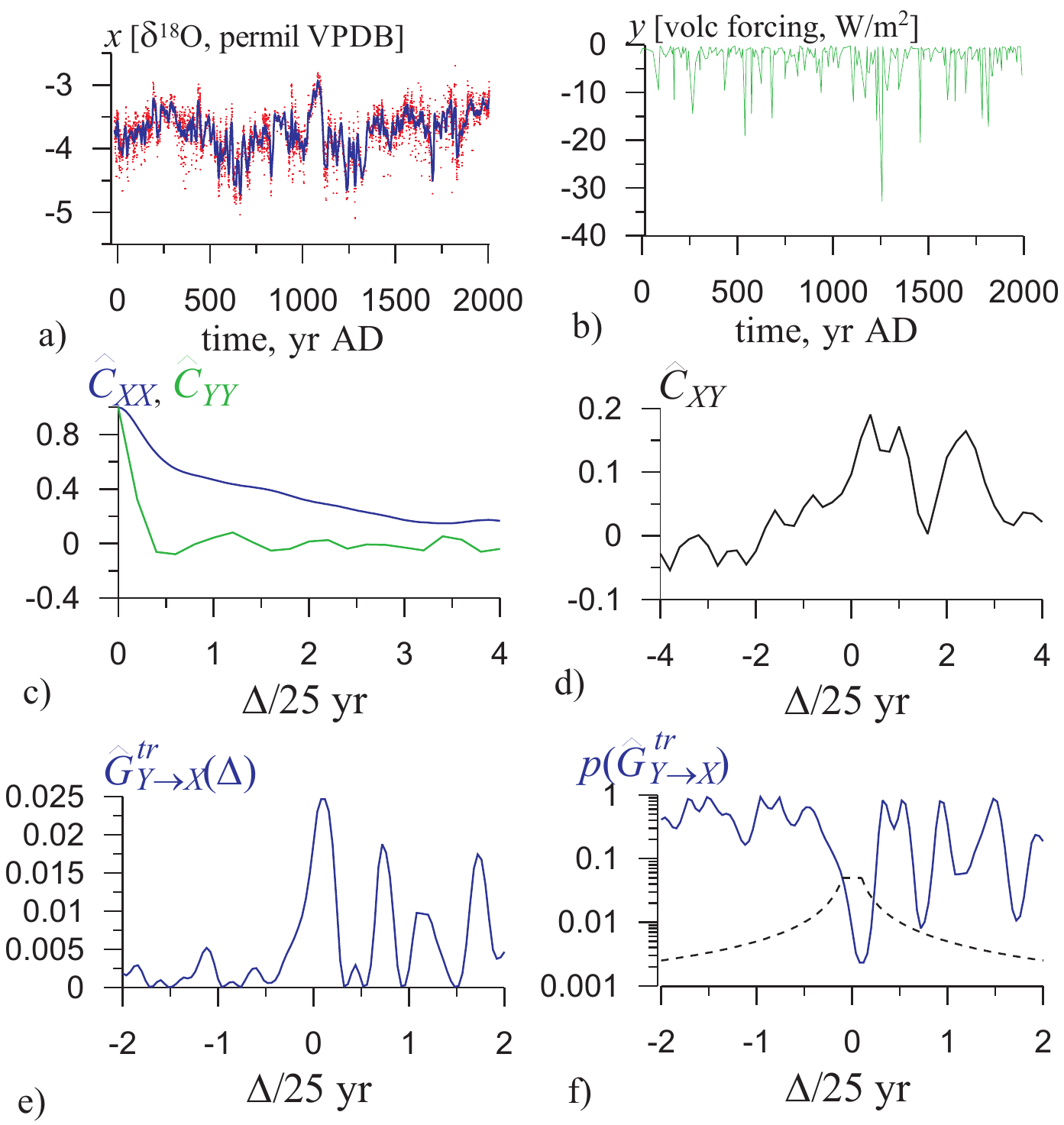}
\caption{\label{FigureS7} Estimation from palaeoclimate data over
the period [15 yr BC - 2010 yr AD]: (a) time series of
$\delta^{18}O$ from a speleothem representing local climate
(moisture) in the Atlantic region, red points denote the original
data, blue line -- signal which is smoothed with a Gaussian kernel
of the effective width of 5 yrs; (b) an original proxy time series
of volcanic activity (global); (c) sample ACF for the signals $x$
(blue) and $y$ smoothed with a Gaussian kernel of 5 yrs width
(green); (d) sample CCF; (e) truncated WG causalities in the
directions volcanoes $\to$ Belize climate for $l_X = 3$, $l_{XY} =
1$; (f) the respective pointwise $p$-levels for the positivity of
$\hat G^{tr}_{Y \to X}$, black dashed lines show the pointwise
$p$-levels corresponding to the total $p$-level of 0.05.}
\end{figure}

We have performed an additional analysis with quite accurately
dated recently published volcanic activity data~\cite{Sigl2015}
(Fig.~\ref{FigureS7}). We have revealed that the volcanic activity influences
 the $\delta^{18}O$ variations with $\Delta = 2-3$ yrs (Fig.~\ref{FigureS7},e,f)
which corresponds well with the maximum point $\Delta = h/2$
(which would be equal to 2.5 yrs here) expected for a non-delayed
coupling and absent dating errors. The deviation is less than 1
yr. The obtained WG causality estimate is statistically highly
significant and the observed small time lag is perfectly
acceptable.

Considering quite precise dating of the volcanic activity proxy,
the above result is a strong argument in favor of an accurate
dating of the speleothem data as well. Since we have found a
``non-physical'' negative lag of total solar irradiance (TSI)
variations behind the speleothem-based hydroclimate proxy, we
suggest that it is the solar activity signal which might be less
accurately dated (with a possible 20 yrs age underestimation, i.e.
the TSI record might be in sections too young) rather than the
speleothem-based hydroclimate proxy. This notion is corroborated
by previous studies, e.g. Ref.~\cite{Steinhilber2012}
(Supplementary material, Figure caption) where the authors also
found 22 yrs negative lag (the TSI impossibly following Asian
monsoon) and concluded that to be acceptable and within errors.

\end{document}